\documentclass[preprint2]{proto}
\usepackage{times}

\newcommand{\refs}{\par\noindent\hangindent=1pc\hangafter=1}

\newcommand{\HeI}{He {\sc i}\relax}%
\newcommand{\CaI}{Ca {\sc i}\relax}%
\newcommand{\TiI}{Ti {\sc i}\relax}%

\newcommand{\MeanB}{$\bar B$}
\newcommand{\Bz}{$B_z$}
\begin{document}

\title{\textbf{\LARGE Magnetospheric Accretion in Classical T Tauri Stars}}

\author{\textbf{\large J. Bouvier, S.~H.~P. Alencar}}
\affil{\small\em Laboratoire d'Astrophysique Grenoble}
\author{\textbf{\large T.~J. Harries}}
\affil{\small\em University of Exeter}
\author{\textbf{\large C.~M. Johns-Krull}}
\affil{\small\em Rice University}
\author{\textbf{\large M.~M. Romanova}}
\affil{\small\em Cornell University}

\begin{abstract}
\baselineskip = 11pt
\leftskip = 0.65in 
\rightskip = 0.65in
\parindent=1pc
%\begin{abstract}
%\begin{list}{ } {\rightmargin 1in}
%{\leftmargin 0in}
%\baselineskip = 11pt
%rule{4.75in}{0.5pt}
%\vskip 1pt
\parindent=1pc {\small The inner 0.1 AU around accreting T Tauri stars
hold clues to many physical processes that characterize the early
evolution of solar-type stars. The accretion-ejection connection takes
place at least in part in this compact magnetized region around the
central star, with the inner disk edge interacting with the star's
magnetosphere thus leading simultaneously to magnetically channeled
accretion flows and to high velocity winds and outflows. The magnetic
star-disk interaction is thought to have strong implications for the
angular momentum evolution of the central system, the inner structure
of the disk, and possibly for halting the migration of young planets
close to the stellar surface. We review here the current status of
magnetic field measurements in T Tauri stars, the recent modeling
efforts of the magnetospheric accretion process, including both
radiative transfer and multi-D numerical simulations, and summarize
current evidence supporting the concept of magnetically-channeled
accretion in young stars. We also discuss the limits of the models and
highlight observational results which suggest that the star-disk
interaction is a highly dynamical and time variable process in young
stars.  \\~\\~\\~}

%\end{list}
\end{abstract}  

\section{\textbf{THE MAGNETIC ACCRETION PARADIGM}}
\bigskip

T Tauri stars are low-mass stars with an age of a few million years, still
contracting down their Hayashi tracks towards the main sequence. Many of
them, the so-called classical T Tauri stars (CTTSs), show signs of
accretion from a circumstellar disk (see, e.g., {\em M\'enard and Bertout},
1999 for a review). Understanding the accretion process in T Tauri stars is
one of the major challenges in the study of pre-main sequence evolution.
Indeed, accretion has a significant and long lasting impact on the
evolution of low mass stars by providing both mass and angular momentum.
The evolution and ultimate fate of circumstellar accretion disks have also
become increasingly important issues since the discovery of extrasolar
planets and planetary systems with unexpected properties.  Deriving the
properties of young stellar systems, of their associated disks and outflows
is therefore an important step towards the establishment of plausible
scenarios for star and planet formation.

The general paradigm of magnetically controlled accretion onto a compact
object is used to explain many of the most fascinating objects in the
Universe. This model is a seminal feature of low mass star formation, but
it is also encountered in theories explaining accretion onto white dwarf
stars (the AM Her stars, e.g., {\em Warner}, 2004), accretion onto pulsars
(the pulsating X-ray sources , e.g., {\em Ghosh and Lamb}, 1979a), and
accretion onto black holes at the center of AGNs and microquasars ({\em
  Koide et al.}, 1999).  Strong surface magnetic fields have long been
suspected to exist in TTSs based on their powerful X-ray and centrimetric
radio emissions ({\em Montmerle et al.}, 1983; {\em Andr\'e}, 1987).
Surface fields of order of 1-3 kG have recently been derived from Zeeman
broadening measurements of CTTS photospheric lines ({\em Johns Krull et
  al.}, 1999a, 2001; {\em Guenther et al.}, 1999) and from the detection of
electron cyclotron maser emission ({\em Smith et al.}, 2003). These strong
stellar magnetic fields are believed to significantly alter the accretion
flow in the circumstellar disk close to the central star.

\begin{figure}[h]
\epsscale{1.0} \plotone{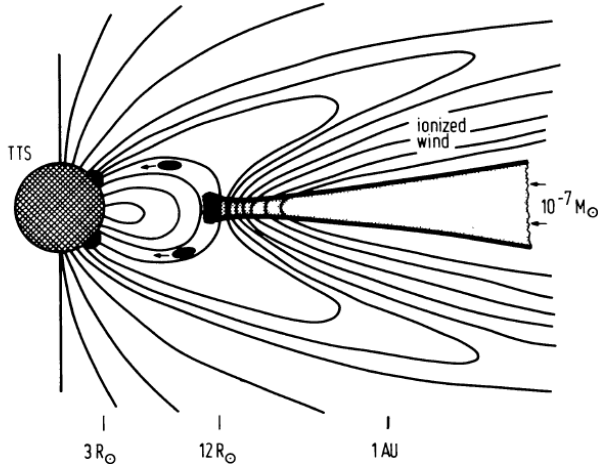}
\caption{A sketch of the basic concept of magnetospheric
        accretion in T Tauri stars (from {\em Camenzind}, 1990).}
\label{cam}
\end{figure}

Based on models originally developed for magnetized compact objects in
X-ray pulsars ({\em Ghosh and Lamb}, 1979a) and {\it assuming\/} that T
Tauri magnetospheres are predominantly dipolar on the large scale,
{\em Camenzind} (1990) and {\em K\"onigl} (1991) showed that the inner
accretion disk is expected to be truncated by the magnetosphere at a
distance of a few stellar radii above the stellar surface for typical
mass accretion rates of 10$^{-9}$ to 10$^{-7}$ M$_\odot$ yr$^{-1}$ in
the disk ({\em Basri and Bertout}, 1989; {\em Hartigan et al.}, 1995;
{\em Gullbring et al.}, 1998). Disk material is then channeled from
the disk inner edge onto the star along the magnetic field lines, thus
giving rise to magnetospheric accretion columns. As the free falling
material in the funnel flow eventually hits the stellar surface,
accretion shocks develop near the magnetic poles. The basic concept of
magnetospheric accretion in T Tauri stars is illustrated in
Figure~\ref{cam}.

The successes and limits of current magnetospheric accretion models in
accounting for the observed properties of classical T Tauri systems are
reviewed in the next sections. Sect. 2 summarizes the current status of
magnetic field measurements in young stars, Sect. 3 provides an account
of current radiative transfer models developed to reproduce the observed
line profiles thought to form at least in part in accretion funnel flows,
Sect. 4 reviews current observational evidence for a highly dynamical
magnetospheric accretion process in CTTSs, and Sect. 5 describes the most
recent 2D and 3D numerical simulations of time dependent star-disk magnetic
interaction.

\bigskip

\section{\textbf{MAGNETIC FIELD MEASUREMENTS}}

\bigskip

\subsection{Theoretical Expectations for T Tauri Magnetic Fields}

While the interaction of a stellar magnetic field with an accretion disk is
potentially very complicated (e.g., {\em Ghosh and Lamb}, 1979a,b), we
present here some results from the leading treatments applied to young
stars.

The theoretical idea behind magnetospheric accretion is that the ram
pressure of the accreting material ($P_{ram} = 0.5 \rho v^2$) will at
some point be offset by the magnetic pressure ($P_B = B^2/8\pi$) for a
sufficiently strong stellar field.  Where these two pressures are
equal, if the accreting material is sufficiently ionized, its motion
will start to be controlled by the stellar field.  This point is
usually referred to as the truncation radius ($R_T$).  If we consider
the case of spherical accretion, the magnetic field becomes
$$B^2 = {\dot M v \over r^2}.\eqno(2.1)$$ 
If we then assume a dipolar stellar magnetic field where $B = B_*(R_*/r)^3$
and set the velocity of the accreting material equal to the free-fall
speed, the radius at which the magnetic field pressure balances the 
ram pressure of the accreting material is
$${R_T \over R_*} = {B_*^{4/7} R_*^{5/7} \over \dot M^{2/7} (2 G M_*)^{1/7}}
= 7.1 B_3^{4/7} \dot M_{-8}^{-2/7} M_{0.5}^{-1/7} R_2^{5/7},\eqno(2.2)$$
where $B_3$ is the stellar field strength in kG, $\dot M_{-8}$ is the mass
accretion rate in units of $10^{-8}$ M$_\odot$ yr$^{-1}$, $M_{0.5}$ is the
stellar mass in units of 0.5 M$_\odot$, and $R_2$ is the stellar radius in
units of $2~{\rm R}_\odot$.  Then, for $B_* = 1$ kG and typical CTTS
properties ($M_* = 0.5$~M$_\odot$, $R_* = 2~{\rm R}_\odot$, and $\dot M =
10^{-8}~{\rm M}_\odot \, {\rm yr}^{-1}$), the truncation radius is about
7 stellar radii.

In the case of disk accretion, the coefficient above is changed, but
the scaling with the stellar and accretion parameters remains the
same.  In accretion disks around young stars, the radial motion due to
accretion is relatively low while the Keplerian velocity due to the
orbital motion is only a factor of $2^{1/2}$ lower than the free-fall
velocity.  The low radial velocity of the disk means that the disk
densities are much higher than in the spherical case, so that the disk
ram pressure is higher than the ram pressure due to spherical
free-fall accretion.  As a result, the truncation radius will move
closer to the star.  In this regard, equation 2.2 gives an upper limit
for the truncation radius.  As we will discuss below, this may be
problematic when we consider the current observations of stellar
magnetic fields.  In the case of disk accretion, another important
point in the disk is the corotation radius, $R_{CO}$, where the
Keplerian angular velocity is equal to the stellar angular velocity.
Stellar field lines which couple to the disk outside of $R_{CO}$ will
act to slow the rotation of the star down, while field lines which
couple to the disk inside $R_{CO}$ will act to spin the star up.
Thus, the value of $R_T$ relative to $R_{CO}$ is an important quantity
in determining whether the star speeds up or slows down its rotation.
For accretion onto the star to proceed, we have the relation $R_T <
R_{CO}$.  This follows from the idea that at the truncation radius and
interior to that, the disk material will be locked to the stellar
field lines and will move at the same angular velocity as the star.
Outside $R_{CO}$ the stellar angular velocity is greater than the
Keplerian velocity, so that any material there which becomes locked to
the stellar field will experience a centrifugal force that tries to
fling the material away from the star.  Only inside $R_{CO}$ will the
net force allow the material to accrete onto the star.

Traditional magnetospheric accretion theories as applied to stars
(young stellar objects, white dwarfs, and pulsars) suggest that the
rotation rate of the central star will be set by the Keplerian
rotation rate in the disk near the point where the disk is truncated
by the stellar magnetic field when the system is in equilibrium.
Hence these theories are often referred to as disk locking theories.
For CTTSs, we have a unique opportunity to test these theories since
all the variables of the problem (stellar mass, radius, rotation rate,
magnetic field, and disk accretion rate) are measureable in principle
(see {\em Johns--Krull and Gafford}, 2002).  Under the assumption that
an equilibrium situation exists, {\em K\"onigl} (1991), {\em Cameron
and Campbell} (1993), and {\em Shu et al.} (1994) have all
analytically examined the interaction between a dipolar stellar
magnetic field (aligned with the stellar rotation axis) and the
surrounding accretion disk.  As detailed in {\em Johns--Krull et al.}
(1999b), one can solve for the surface magnetic field strength on a
CTTS implied by each of these theories given the stellar mass,
radius, rotation period, and accretion rate.  For the work of {\em
K\"onigl} (1991), the resulting equation is:
$$
B_* = 3.43
\Bigl({\epsilon \over 0.35}\Bigr)^{7/6} 
\Bigl({\beta \over 0.5}\Bigr)^{-7/4}
\Bigl({M_* \over M_\odot}\Bigr)^{5/6}\times$$
$$
\times \Bigl({\dot{M} \over 10^{-7}\; M_\odot\; {\rm yr}^{-1}}\Bigr)^{1/2}
\Bigl({R_* \over R_\odot}\Bigr)^{-3}
\Bigl({P_* \over 1\; {\rm dy}}\Bigr)^{7/6} \,\, {\rm kG},
\eqno(2.3)
$$ In the work of {\em Cameron and Campbell} (1993) the equation for
the stellar field is:
$$
B_* = 1.10
\gamma^{-1/3}
\Bigl({M_* \over M_\odot}\Bigr)^{2/3}
\Bigl({\dot{M} \over 10^{-7}\; M_\odot\; {\rm yr}^{-1}}\Bigr)^{23/40}\times$$
$$
\times\Bigl({R_* \over R_\odot}\Bigr)^{-3}
\Bigl({P_* \over 1\; {\rm dy}}\Bigr)^{29/24} \,\, {\rm kG},
\eqno(2.4)
$$
Finally, from {\em Shu et al.} (1994), the resulting equation is:
$$
B_*  = 3.38
\Bigl({\alpha_x \over 0.923}\Bigr)^{-7/4}
\Bigl({M_* \over M_\odot}\Bigr)^{5/6}
\Bigl({\dot{M} \over 10^{-7}\; M_\odot\; {\rm yr}^{-1}}\Bigr)^{1/2}\times$$
$$
\times\Bigl({R_* \over R_\odot}\Bigr)^{-3}
\Bigl({P_* \over 1\; {\rm dy}}\Bigr)^{7/6} \,\, {\rm kG},
\eqno(2.5)
$$ All these equations contain uncertain scaling parameters
($\epsilon, \beta, \gamma, \alpha_x$) which characterize the
efficiency with which the stellar field couples to the disk or the
level of vertical shear in the disk.  Each study presents a best
estimate for these parameters allowing the stellar field to be
estimated (Table~1).  Observations of magnetic fields on CTTSs can then
serve as a test of these models.

     To predict magnetic field strengths for specific CTTSs, we need
observational estimates for certain system parameters.  We adopt
rotation periods from {\em Bouvier et al.} (1993, 1995) and stellar
masses, radii, and mass accretion rates from {\em Gullbring et al.}
(1998).  Predictions for each analytic study are presented in Table~1.
Note, these field strengths are the equatorial values.  The field at
the pole will be twice these values and the average over the star will
depend on the exact inclination of the dipole to the observer, but for
$i = 45^\circ$ the average field strength on the star is $\sim 1.4$
times the values given in the Table.  Because of differences in
underlying assumptions, these predictions are not identical, but they
do have the same general dependence on system characteristics.
Consequently, field strengths predicted by the 3 theories, while
different in scale, nonetheless have the same pattern from star to
star.  Relatively weak fields are predicted for some stars (DN Tau, IP
Tau), but detectably strong fields are expected on stars such as BP
Tau.

\begin{deluxetable}{lccccccccc}
\tablewidth{15.0truecm}
\tablecaption{Predicted Magnetic Field Strengths}
\tablehead{
   \colhead{}&
   \colhead{$M_*$}&
   \colhead{$R_*$}&
   \colhead{$\dot{M} \times 10^8$}&
   \colhead{$P_{rot}$}&
   \colhead{$B^a_*$}&
   \colhead{$B^b_*$}&
   \colhead{$B^c_*$}&
   \colhead{$R_{CO}$}&
   \colhead{$\bar B_{obs}$}\\[0.2ex]
%   \colhead{$R_T$}
   \colhead{Star}&
   \colhead{$(M_\odot)$}&
   \colhead{$(R_\odot)$}&
   \colhead{$(M_\odot {\rm yr}^{-1})$}&
   \colhead{(days)}&
   \colhead{(G)}&
   \colhead{(G)}&
   \colhead{(G)}&
   \colhead{($R_*$)}&
   \colhead{(kG)}
%   \colhead{($R_*$)}
} \startdata AA Tau & 0.53 & 1.74 & 0.33 & 8.20 & 810 & 240 & 960 &
8.0 & 2.57 \\ BP Tau & 0.49 & 1.99 & 2.88 & 7.60 & 1370 & 490 & 1620 &
6.4 & 2.17 \\ CY Tau & 0.42 & 1.63 & 0.75 & 7.90 & 1170 & 390 & 1380 &
7.7 & \\ DE Tau & 0.26 & 2.45 & 2.64 & 7.60 & 420 & 164 & 490 & 4.2 &
1.35 \\ DF Tau & 0.27 & 3.37 & 17.7 & 8.50 & 490 & 220 & 570 & 3.4 &
2.98 \\ DK Tau & 0.43 & 2.49 & 3.79 & 8.40 & 810 & 300 & 950 & 5.3 &
2.58 \\ DN Tau & 0.38 & 2.09 & 0.35 & 6.00 & 250 & 80 & 300 & 4.8 &
2.14 \\ GG Tau A& 0.44 & 2.31 & 1.75 & 10.30 & 890 & 320 & 1050 & 6.6
& 1.57 \\ GI Tau & 0.67 & 1.74 & 0.96 & 7.20 & 1450 & 450 & 1700 & 7.9
& 2.69 \\ GK Tau & 0.46 & 2.15 & 0.64 & 4.65 & 270 & 90 & 320 & 4.2 &
2.13 \\ GM Aur & 0.52 & 1.78 & 0.96 & 12.00 & 1990 & 660 & 2340 & 10.0
& \\ IP Tau & 0.52 & 1.44 & 0.08 & 3.25 & 240 & 60 & 280 & 5.2 & \\ TW
Hya & 0.70 & 1.00 & 0.20 & 2.20 & 900 & 240 & 1060 & 6.3 & 2.61 \\ T
Tau & 2.11 & 3.31 & 4.40 & 2.80 & 390 & 110 & 460 & 3.2 & 2.39 \\
\enddata \tablecomments{Magnetic field values come from applying the
theory of ($a$) {\em K\"onigl} (1991), ($b$) {\em Cameron and
Campbell}, or ($c$) {\em Shu et al.}  (1994). These are the equatorial
field strengths assuming a dipole magnetic field.}
\end{deluxetable}

\subsection{Measurement Techniques}

Virtually all measurements of stellar magnetic fields make use of the
Zeeman effect.  Typically, one of two general aspects of the Zeeman effect
is utilized: (1) Zeeman broadening of magnetically sensitive lines observed
in intensity spectra, or (2) circular polarization of magnetically
sensitive lines.  Due to the nature of the Zeeman effect, the splitting due
to a magnetic field is proportional to $\lambda^2$ of the transition.
Compared with the $\lambda^1$ dependence of Doppler line broadening
mechanisms, this means that observations in the infrared (IR) are generally
more sensitive to the presence of magnetic fields than optical
observations.

%When an atom is in a magnetic field,
%different projections of the total orbital angular momentum are no
%longer degenerate, shifting the energy levels taking part in the
%transition.  In the simple Zeeman effect, a spectral line splits into
%3 components: 2 $\sigma$ components split to either side of the
%nominal line center and 1 unshifted $\pi$ component.  The wavelength
%shift of a given $\sigma$ component is given by
%$$\Delta\lambda = {e \over 4 \pi m_e c^2} \lambda^2 g B, \eqno(2.6)$$
%where $g$ is the Land\'e g-factor of the specific transition, $B$ is the 
%strength of the magnetic field, and $\lambda$ is the wavelength of the 
%transition.  Evaluating the constants, the wavelength shift is
%$$\Delta\lambda = 4.67\times10^{-7}\ \lambda^2 g B\; {\rm m\AA},
%\eqno(2.7)$$ where $\lambda$ is in \AA\ and $B$ is in kG. One thing
%to note from this equation is the $\lambda^2$ dependence of the Zeeman
%effect.  Compared with the $\lambda^1$ dependence of Doppler line
%broadening mechanisms, this means that observations in the infrared
%(IR) are generally more sensitive to the presence of magnetic fields
%than optical observations.

The simplest model of the spectrum from a magnetic star assumes that
the observed line profile can be expressed as $F(\lambda) =
F_B(\lambda)*f + F_Q(\lambda)*(1-f)$; where $F_B$ is the spectrum
formed in magnetic regions, $F_Q$ is the spectrum formed in
non-magnetic (quiet) regions, and $f$ is the flux weighted surface
filling factor of magnetic regions.  The magnetic spectrum, $F_B$,
differs from the spectrum in the quiet region not only due to Zeeman
broadening of the line, but also because magnetic fields affect
atmospheric structure, causing changes in both line strength and
continuum intensity at the surface.  Most studies {\it assume} that
the magnetic atmosphere is in fact the same as the quiet atmosphere
because there is no theory to predict the structure of the magnetic
atmosphere.  If the stellar magnetic field is very strong, the
splitting of the $\sigma$ components is a substantial fraction of the
line width, and it is easy to see the $\sigma$ components sticking out
on either side of a magnetically sensitive line.  In this case, it is
relatively straightforward to measure the magnetic field strength,
$B$.  Differences in the atmospheres of the magnetic and quiet regions
primarily affect the value of $f$.  If the splitting is a small
fraction of the intrinsic line width, then the resulting observed
profile is only subtly different from the profile produced by a star
with no magnetic field and more complicated modelling is required to
be sure all possible non-magnetic sources (e.g., rotation and pressure
broadening) have been properly constrained.

     In cases where the Zeeman broadening is too subtle to detect
directly, it is still possible to diagnose the presence of magnetic
fields through their effect on the equivalent width of magnetically
sensitive lines.  For strong lines, the Zeeman effect moves the
$\sigma$ components out of the partially saturated core into the line
wings where they can effectively add opacity to the line and increase
the equivalent width.  The exact amount of equivalent width increase
is a complicated function of the line strength and Zeeman splitting
pattern ({\em Basri et al.}, 1992).  This method is primarily
sensitive to the product of $B$ multiplied by the filling factor $f$
({\em Basri et al.}, 1992, {\em Guenther et al.}, 1999).  Since this
method relies on relatively small changes in the line equivalent
width, it is very important to be sure other atmospheric parameters
which affect equivalent width (particularly temperature) are
accurately measured.

     Measuring circular polarization in magnetically sensitive lines
is perhaps the most direct means of detecting magnetic fields on
stellar surfaces, but is also subject to several limitations.  When
viewed along the axis of a magnetic field, the Zeeman $\sigma$
components are circularly polarized, but with opposite helicity; and
the $\pi$ component is absent.  The helicity of the $\sigma$
components reverses as the polarity of the field reverses.  Thus, on a
star like the Sun that typically displays equal amounts of $+$ and $-$
polarity fields on its surface, the net polarization is very small.
If one magnetic polarity does dominate the visible surface of the
star, net circular polarization is present in Zeeman sensitive lines,
resulting in a wavelength shift between the line observed through
right- and left-circular polarizers.  The magnitude of the shift
represents the surface averaged line of sight component of the
magnetic field (which on the Sun is typically less than 4 G even
though individual magnetic elements on the solar surface range from
$\sim 1.5$ kG in plages to $\sim 3.0$ kG in spots).  Several
polarimetric studies of cool stars have generally failed to detect
circular polarization, placing limits on the disk-averaged magnetic
field strength present of $10-100$ G (e.g., {\em Vogt}, 1980; {\em
Brown and Landstreet}, 1981; {\em Borra et al.}, 1984).  One notable
exception is the detection of circular polarization in segments of the
line profile observed on rapidly rotating dwarfs and RS CVn stars
where Doppler broadening of the line ``resolves" several independent
strips on the stellar surface (e.g., {\em Donati et al.}, 1997; {\em
Petit et al.}, 2004; {\em Jardine et al.}, 2002).

\subsection{Mean Magnetic Field Strength}

TTS typically have $v\sin i$ values of 10 km s$^{-1}$, which means
that observations in the optical typically cannot detect the actual
Zeeman broadening of magnetically sensitive lines because the
rotational broadening is too strong.  Nevertheless, optical
observations can be used with the equivalent width technique to detect
stellar fields.  {\em Basri et al.} (1992) were the first to detect a
magnetic field on the surface of a TTS, inferring a value of $Bf =
1.0$ kG on the NTTS Tap 35.  For the NTTS Tap 10, {\em Basri et al.}
(1992) find only an upper limit of $Bf < 0.7$ kG.  {\em Guenther et
al.} (1999) apply the same technique to spectra of 5 TTSs, claiming
significant field detections on two stars; however, these authors
analyze their data using models off by several hundred K from the
expected effective temperature of their target stars, always a concern
when relying on equivalent widths.

As we saw above, observations in the IR will help solve the difficulty in
detecting direct Zeeman broadening. For this reason and given the
temperature of most TTSs (K7 - M2), Zeeman broadening measurements for these
stars are best done using several \TiI\ lines found in the K band.  Robust
Zeeman broadening measurements require Zeeman insensitive lines to
constrain nonmagnetic broadening mechanisms.  Numerous CO lines at 2.31
$\mu$m have negligible Land\'e-$g$ factors, making them an ideal null
reference.

It has now been shown that the Zeeman insensitive CO lines are well
fitted by models with the same level of rotational broadening as that
determined from optical line profiles ({\em Johns--Krull and Valenti},
2001; {\em Johns--Krull et al.}, 2004; {\em Yang et al.}, 2005).  In
contrast, the 2.2 $\mu$m \TiI\ lines cannot be fitted by models
without a magnetic field.  Instead, the observed spectrum is best fit
by a model with a superposition of synthetic spectra representing
different regions on the star with different magnetic field strengths.
Typically, the field strengths in these regions are assumed to have
values of 0, 2, 4, and 6 kG and only the filling factor of each region
is solved for.  The resulting magnetic field distribution is unique
because the Zeeman splitting produced by a 2 kG field is comparable to
the nonmagnetic width of the \TiI\ spectral lines. In other words, the
Zeeman resolution of the \TiI\ lines is about 2 kG (see {\em
Johns--Krull et al.}, 1999b, 2004).

The intensity-weighted mean magnetic field strength, \MeanB, over the
entire surface of most TTSs analyzed to date is $\sim 2.5$ kG, with field
strengths reaching at least 4 kG and probably even 6 kG in some regions.
Thus, magnetic fields on TTSs are stronger than on the Sun, even though the
surface gravity on these stars is lower by a factor of ten. On the Sun and
other main-sequence stars, magnetic field strength seems to be set by an
equipartition of gas and magnetic pressure. In contrast, the photospheres
of TTSs are apparently dominated by magnetic pressure, rather than gas
pressure (see also {\em Johns--Krull et al.}, 2004).  Strong magnetic
fields are ubiquitous on TTSs. By fitting IR spectra, magnetic field
distributions for several TTSs have now been measured ({\em Johns--Krull et
  al.}, 1999b, 2001, 2004; {\em Yang et al.}, 2005).  Many of these field
strengths are reported in Table~1.

\subsection{Magnetic Field Topology}

Zeeman broadening measurements are sensitive to the distribution of
magnetic field strengths, but they have limited sensitivity to
magnetic geometry. In contrast, circular polarization measurements for
individual spectral lines are sensitive to magnetic geometry, but they
provide limited information about field strength. The two techniques
complement each other well, as we demonstrate below.

Most existing magnetospheric accretion models assume that intrinsic TTS
magnetic fields are dipolar, but this would be unprecedented for cool
stars.  The higher order components of a realistic multi-polar field will
fall off more rapidly with distance than the dipole component, so at the
inner edge of the disk a few stellar radii from the surface, it is likely
the dipolar component of the stellar field will dominate. However, at the
stellar surface the magnetic field is likely to be more complicated.  In
support of this picture is the fact that spectropolarimetric observations
do not detect polarization in photospheric absorption lines: {\em Brown and
  Landstreet} (1981) failed to detect polarization in T Tau and two FU Ori
objects; {\em Johnstone and Penston} (1986) observed 3 CTTSs and reported a
marginal field detection for RU Lup, but they were not able to confirm the
signal in a subsequent observation, perhaps because of rotational
modulation ({\em Johnstone and Penston}, 1987); {\em Donati et al.} (1997)
find no evidence for a strong dipolar field component in the 3 TTSs they
observed; {\em Johns--Krull et al.} (1999a) failed to detect polarization
in the photospheric lines of BP Tau; and {\em Valenti and Johns--Krull}
(2004) do not detect significant polarization in the photospheric lines of
4 CTTSs each observed over a rotation period.  {\em Smirnov et al.} (2003)
report a marginal detection of circular polarization in the lines of T Tau
corresponding to a field of $\sim 150 \pm 50$ G; however, {\em Smirnov et
  al.}  (2004) and {\em Daou et al.} (2005) failed to confirm this
detection, placing an upper limit on the field of $\leq 120$ G for T Tau.

However, {\em Johns--Krull et al.} (1999a) did discover circular
polarization in CTTS emission line diagnostics that form predominantly
in the accretion shock at the surface of the star. This circular
polarization signal is strongest in the narrow component of the \HeI\
5876 \AA\ emission line, but it is also present in the \CaI\ infrared
triplet lines. The peak value of \Bz\ is 2.5 kG, which is comparable
to our measured values of \MeanB.  Circular polarization in the \HeI\
5876 \AA\ emission line has now been observed in a number of CTTSs
({\em Valenti et al.}, 2004; {\em Symington et al.}, 2005b).  Note,
since this polarization is detected in a line associated with the
accretion shock on CTTSs, it forms over an area covering typically $<
5$\% of the stellar surface ({\em Valenti et al.}, 1993; {\em Calvet
and Gullbring}, 1998).  While the field in this 5\% of the star appears
to be highly organized (and as discussed below may trace the dipole
component of the field at the surface), the lack of polarization
detected in photospheric lines forming over the entire surface of the
star strongly rule out a global dipole geometry for the entire field.

\begin{figure}[t]
\epsscale{1.0} \plotone{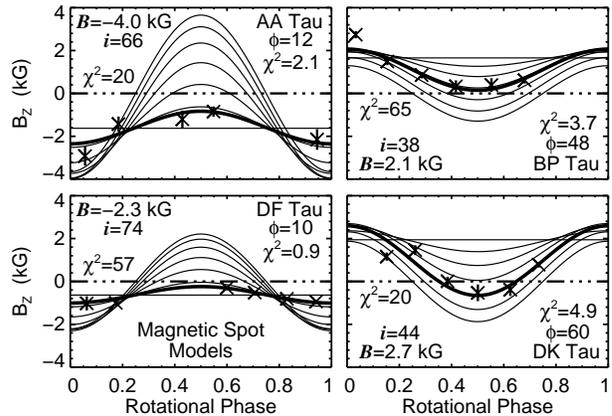}
\caption{Variations in the circular polarization of the \HeI\ emission
line as a function of rotation phase for 4 CTTSs.  Polarization levels
are translated into \Bz\ values in the line formation region.  Vertical
bars centered on each measurement ($\times$) give the 1$\sigma$
uncertainty in the field measurement.  Solid lines show predicted
rotational modulation in \Bz\ for a single magnetic spot at latitudes
($\phi$) ranging from 0$^\circ$ to 90$^\circ$ in 15$^\circ$ increments.
The best fit latitude is shown in the thick solid curve.}
\label{cjk_fig1}
\end{figure}

Figure~\ref{cjk_fig1} shows measurements of \Bz\ on 6 consecutive nights.
These measurements were obtained at McDonald Observatory, using the Zeeman
analyzer described by {\em Johns--Krull et al.} (1999a).  The measured
values of \Bz\ vary smoothly on rotational timescales, suggesting that
uniformly oriented magnetic field lines in accretion regions sweep out a
cone in the sky, as the star rotates. Rotational modulation implies a lack
of symmetry about the rotation axis in the accretion or the magnetic field
or both. For example, the inner edge of the disk could have a concentration
of gas that corotates with the star, preferentially illuminating one sector
of a symmetric magnetosphere. Alternatively, a single large scale magnetic
loop could draw material from just one sector of a symmetric disk. 

Figure~\ref{cjk_fig1} shows one interpretation of the \HeI\
polarization data.  Predicted values of \Bz\ are shown for a simple
model consisting of a single magnetic spot at latitude $\phi$ that
rotates with the star. The magnetic field is assumed to be radial with
a strength equal to our measured values of \MeanB.  Inclination of the
rotation axis is constrained by measured $v\sin i$ and rotation
period, except that inclination ($i$) is allowed to float when it
exceeds $60^\circ$ because $v\sin i$ measurements cannot distinguish
between these possibilities.  Predicted variations in \Bz\ are plotted
for spot latitudes ranging from $0^\circ$ to $90^\circ$ in $15^\circ$
increments. The best fitting model is shown by the thick curve. The
corresponding spot latitude and reduced $\chi^2$ are given on the
right side of each panel. The null hypothesis (that no polarization
signal is present) produces very large values of $\chi^2$ which are
given on the left side of each panel.  In all four cases, this simple
magnetic spot model reproduces the observed \Bz\ time series.  The
\HeI\ rotationally modulated polarization combined with the lack of
detectable polarization in photospheric absorption lines as described
above paints a picture in which the magnetic field on TTSs displays a
complicated geometry at the surface which gives way to a more ordered,
dipole-like geometry a few stellar radii from the surface where the
field intersects the disk.  The complicated surface topology results
in no net polarization in photospheric absorption lines, but the
dipole-like geometry of the field at the inner disk edge means that
accreting material follows these field lines down to the surface so
that emission lines formed in the accretion shock preferentially
illuminate the dipole component of the field, producing substantial
circular polarization in these emission lines.

\begin{figure}[t]
\includegraphics[scale=0.37,angle=90]{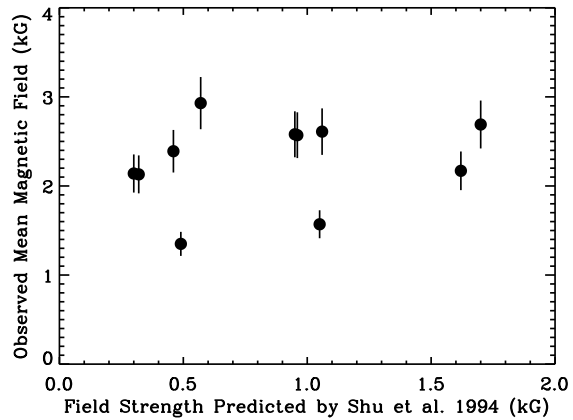} 
%\epsscale{0.9} \plotone{fig_2.2.eps}
\caption{Observed mean magnetic field strength determined from IR
Zeeman broadening measurements as a function of the predicted field
strength from Table 1 for the theory of {\em Shu et al.} (1994).  No
statistically significant correlation is found between the observed
and predicted field strengths. }
\label{cjk_fig2}
\end{figure}

\subsection{Confronting Theory with Observations}

At first glance, it might appear that magnetic field measurements on
TTS are generally in good agreement with theoretical expectations.
Indeed, the IR Zeeman broadening measurements indicate mean fields on
several TTSs of $\sim 2$ kG, similar in value to those predicted in
Table~1 (recall the field values in the Table are the equatorial
values for a dipolar field, and that the mean field is about 1.5 times
these equatorial values).  However, in detail the field observations
do not agree with the theory.  This can be seen in
Figure~\ref{cjk_fig2}, where we plot the measured magnetic field
strengths versus the predicted field strengths from {\em Shu et al.}
(1994, see Table~1).  Clearly, the measured field strengths show no
correlation with the predicted field strengths.  The field topology
measurements give some indication to why there may be a lack of
correlation: the magnetic field on TTSs are not dipolar, and the dipole
component to the field is likely to be a factor of $\sim 10$ or more
lower than the values predicted in Table~1.  As discussed in {\em
Johns--Krull et al.} (1999b), the 3 studies which produce the field
predictions in Table~1 involve uncertain constants which describe the
efficiency with which stellar field lines couple to the accretion
disk.  If these factors are much different than estimated, it may be
that the required dipole components to the field are substantially
less than the values given in the Table.  On the other hand, equation
2.2 was derived assuming perfect coupling of the field and the matter,
so it serves as a firm upper limit to $R_T$ as discussed in \S 2.1.
Spectropolarimetry of TTSs indicates that the dipole component of the
magnetic field is $\leq 0.1$ kG ({\em Valenti and Johns--Krull}, 2004;
{\em Smirnov et al.}, 2004; {\em Daou et al.}, 2005).  Putting this
value into equation 2.2, we find $R_T \leq 1.9~R_*$ for typical CTTS
parameters.  Such a low value for the truncation radius is
incompatible with rotation periods of 7-10 days as found for many CTTSs
(Table~1 and, e.g., {\em Herbst et al.}, 2002).

Does this then mean that magnetospheric accretion does not work?
Independent of the coupling efficiency between the stellar field and
the disk, magnetospheric accretion models predict correlations between
stellar and accretion parameters.  As shown in \S 2.3, the fields on
TTS are found to all be rather uniform in strength.  Eliminating the
stellar field then, {\em Johns--Krull and Gafford} (2002) looked for
correlation among the stellar and accretion parameters, finding little
evidence for the predicted correlations.  This absence of the expected
correlations had been noted earlier by {\em Muzerolle et al.}  (2001).
On the other hand, {\em Johns--Krull and Gafford} (2002) showed how
the models of {\em Ostriker and Shu} (1995) could be extended to take
into account non-dipole field geometries.  Once this is done, the
current data do reveal the predicted correlations, suggesting
magnetospheric accretion theory is basically correct as currently
formulated.  So then, how do we reconcile the current field
measurements with this picture?  While the dipole component of the
field is small on TTSs, it is clear the stars posses strong fields over
most, if not all, of their surface but with a complicated surface
topology.  Perhaps this can lead to a strong enough field so that $R_T
\sim 6~R_*$ as generally suggested by observations of CTTS phenomena.
More complicated numerical modelling of the interaction of a complex
geometry field with an accretion disk will be required to see if this
is feasible.

\bigskip

\section{\textbf{SPECTRAL DIAGNOSTICS OF MA\-GNE\-TO\-SPHE\-RIC ACCRETION}}
\bigskip

Permitted emission line profiles from CTTSs, in particular the Balmer
series, show a wide variety of morphologies including symmetric,
double-peaked, P~Cygni, and inverse P~Cygni (IPC) type ({\em Edwards
et al.}, 1994): common to all shapes is a characteristic line width
indicative of bulk motion within the circumstellar material of
hundreds of km\,s$^{-1}$. The lines themselves encode both geometrical
and physical information on the accretion process and its rate, and
the challenge is to use the profiles to test and refine the
magnetospheric accretion model.

Interpretation of the profiles requires a translational step between
the physical model and the observable spectra; this is the process of
radiative-transfer (RT) modelling. The magnetospheric accretion
paradigm presents a formidable problem in RT, since the geometry is
two or three dimensional, the material is moving, and the
radiation-field and the accreting gas are decoupled (i.e. the problem
is non-LTE).  However, the past decade has seen the development of
increasingly sophisticated RT models that have been used to model line
profiles (both equivalent width and shape) in order to determine accretion rates. In
this section we describe the development of these models, and
characterize their successes and failures.

Current models are based on idealized axisymmetric geometry, in which
the circumstellar density structure is calculated assuming free-fall
along dipolar field lines that emerge from a geometrically thin disc
at a range of radii encompassing the corotation radius. It is assumed
that the kinetic energy of the accreting material is completely
thermalized, and that the accretion luminosity, combined with the area
of the accretion footprints (rings) on the stellar surface, provide
the temperature of the hot spots.  The circumstellar density and
velocity structure is then fully described by the mass accretion rate,
and the outer and inner radii of the magnetosphere in terms of the
photospheric radius of the star ({\em Hartmann et al.}, 1994).

A significant, but poorly constrained, input parameter for the models
is the temperature structure of the accretion flow. This is a
potential pitfall, as the form of the temperature structure may have a
significant impact on the line source functions, and therefore the
line profiles themselves. Self-consistent radiative equilibrium models
({\em Martin}, 1996) indicate that adiabatic heating and cooling via
bremsstrahlung dominate the thermal budget, whereas Hartmann and
co-workers adopt a simple volumetric heating rate combined with a
schematic radiative cooling rate which leads to a temperature
structure that goes as the reciprocal of the density. Thus the
temperature is low near the disc, and passes through a maximum (as the
velocity increases and density decreases) before the stream cools
again as it approaches the stellar surface (and the density increases
once more).

With the density, temperature and velocity structure of the accreting
material in place, the level populations of the particular atom under
consideration must be calculated under the constraint of statistical
equilibrium. This calculation is usually performed using the Sobolev
approximation, in which it is assumed that the conditions in the gas do not
vary significantly over a length scale given by
\begin{equation}
l_{\rm S} = v_{therm} / (dv/dr)
\end{equation}
where $v_{therm}$ is the thermal velocity of the gas and $dv/dr$ is the
velocity gradient. Such an approximation is only strictly valid in the
fastest parts of the accretion flow.  Once the level populations have
converged, the line opacities and emissivities are then computed, allowing
the line profile of any particular transition to be calculated.

\begin{figure}[t]
\epsscale{0.9} \plotone{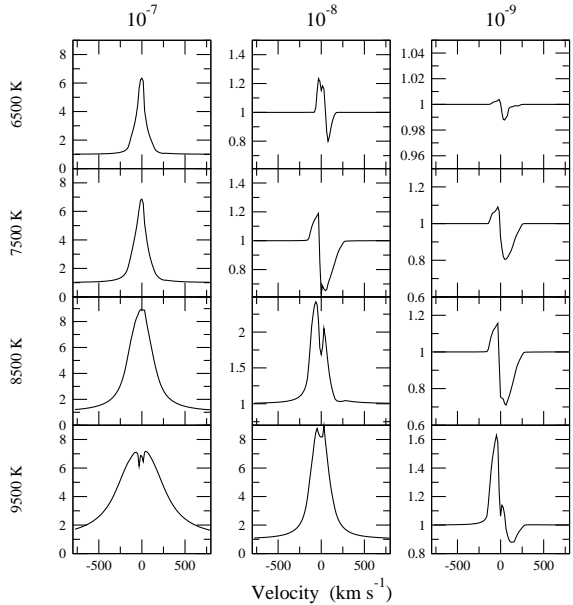}
\caption{H$\alpha$ model profiles for a wide range of mass accretion
rate and accretion flow maximum temperature (from {\em Kurosawa et
al.}, 2006). The profiles are based on canonical CTTS parameters ($R=2~
R_\odot$, $M=0.5~M_\odot$, $T=4000$ K) viewed at an inclination of
$55^\circ$. The maximum temperature of the accretion flow is indicated
along the left of the figure, while the accretion rate (in
$M_\odot$\,yr$^{-1}$) is shown along the top.}
\label{harries_fig1}
\end{figure}

The first models computed using the method outlined above were
presented by {\em Hartmann et al.} (1994), who adopted a two-level
atom approximation.  It was demonstrated that the magnetospheric
accretion model could reproduce the main characteristics of the
profiles, including IPC profiles and blue-shifted central emission
peaks. The original Hartmann et al. model was further improved by {\em
Muzerolle et al.} (2001). Instead of using a two-level
approximation, they solved statistical equilibrium (still under
Sobolev) for a 20-level hydrogen atom. Their line profiles were
computed using a direct integration method, which, unlike the Sobolev
approach, allows the inclusion of Stark broadening effects. It was
found that the broadening was most significant for H$\alpha$, with the
line reaching widths of $\sim 500$ km\,s$^{-1}$, a width that
significantly exceeds the doppler broadening due to infall alone, and
in much better agreement with observation. The H$\beta$ model profiles
were found to be in broad agreement with the observations, in terms of
the velocity of the emission peak ({\em Alencar and Basri}, 2000), and
in the asymmetry of the profiles ({\em Edwards et al.}, 
1994). Figure~\ref{harries_fig1} shows model H$\alpha$ profiles as a
function of mass accretion rate and accretion flow temperature; one
can see that for typical CTTS accretion rates the line profiles are
broadly symmetric although slightly blueshifted -- the reduced optical
depth for the lower accretion rate models yields the IPC morphology.

\begin{figure}[t]
\epsscale{0.7} \plotone{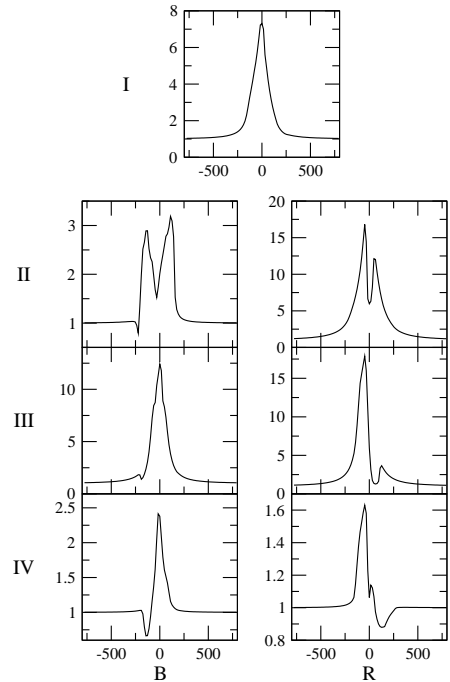}
\caption{Sample H$\alpha$ model profiles ({\em Kurosawa et al.}, 2006) which
  characterize the morphological classification (Types I-IV B/R) by {\em
    Reipurth et al.} (1996). The combination of magnetospheric accretion,
  the accretion disc, and the collimated disk wind can reproduce the wide
  range of H$\alpha$ profiles seen in observations. The horizontal axes are
  velocities in km\,s$^{-1}$ and the vertical axes are continuum normalized
  intensities.}
\label{harries_fig2}
\end{figure}

Axisymmetric models are obviously incapable of reproducing the wide range
of variability that is observed in the emission lines of CTTSs (Sect. 4).
Although the addition of further free parameters to models naturally renders
them more arbitrary, the observational evidence for introducing such
parameters is compelling.  Perhaps the simplest extension is to break the
axisymmetry of the dipole, leaving curtains of accretion in azimuth --
models such as these have been proposed by a number of observers attempting
to explain variability in CTTSs and are observed in MHD simulations ({\em
  Romanova et al.}, 2003). Synthetic time-series for a CTTS magnetosphere
structured along these lines were presented by {\em Symington et al.}
(2005a). It was found that some gross characteristics of the observed line
profiles were produced using a `curtains' model, although the general level
of variability predicted is larger than that observed, suggesting that the
magnetosphere may be characterized by a high degree of axisymmetry, broken
by higher-density streams that produce the variability. 

The emission line profiles of CTTSs often display the signatures of outflow
as well as infall, and recent attempts have been made to account for this
in RT modelling. {\em Alencar et al.} (2005) investigated a dipolar
accretion geometry combined with a disk wind in order to model the line
profile variability of RW~Aur. They discovered that magnetospheric
accretion alone could not simultaneously model H$\alpha$, H$\beta$ and NaD
profiles, and found that the wind contribution to the lines profiles is
quite important in that case.

\begin{figure}[t]
%\epsscale{0.7} 
\includegraphics[scale=0.4,angle=-90]{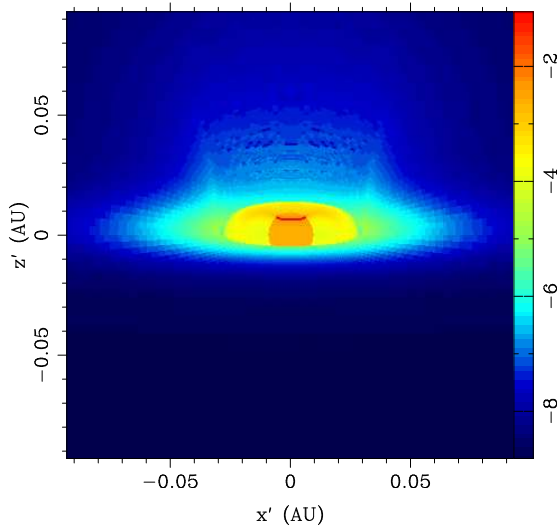} 
%\plotone{harries3.eps}
\caption{A simulated H$\alpha$ image of an accreting CTTS with an
outflow ($\log \dot{M}_{\rm acc} = -8$, $\log \dot{M}_{\rm wind}=-9$)
viewed at an inclination of $80^\circ$. The wind emission is
negligible compared to the emission from the magnetosphere, and the
lower half of the wind is obscured by the circumstellar disk ({\em Kurosawa
et al.}, 2006).}
\label{harries_fig3}
\end{figure}

Hybrid models ({\em Kurosawa et al.}, 2006) combining a standard dipolar
accretion flow with an outflow (e.g., Fig.~\ref{harries_fig3}) are capable
of reproducing the broad range of observed H$\alpha$ profiles
(Fig.~\ref{harries_fig2}). Obviously spectroscopy alone is insufficient to
uniquely identify a set of model parameters for an individual object,
although by combining spectroscopy with other probes of the circumstellar
material, one should be able to reduce the allowable parameter space
considerably. For example linear spectropolarimetry provides a unique
insight into the accretion process; scattering of the line emission by
circumstellar dust imprints a polarization signature on the line which is
geometry dependent. An H$\alpha$ spectropolarimetric survey by {\em Vink et
  al.}  (2005a) revealed that 9 out of 11 CTTSs showed a measurable change
in polarization through the line, while simple numerical models by {\em
  Vink et al.} (2005b) demonstrate that this polarization may be used to
gauge the size of the disk inner hole.

The radiative-transfer models described above are now routinely used to
determine mass accretion rates across the mass spectrum from Herbig AeBe
({\em Muzerolle et al.}, 2004) stars to brown dwarfs ({\em Lawson et al.},
2004; {\em Muzerolle et al.}, 2005), and in the CTTS mass regime at least
the accretion rates derived from RT modelling have been roughly calibrated
against other accretion-rate measures, such as the UV continuum (e.g., {\em
  Muzerolle et al.}, 2001). However, one must be aware of the simplifying
assumptions which underlie the models and that must necessarily impact on
the validity of any quantity derived from them, particularly the mass
accretion rate. Magnetic field measurements (Sect. 2) and time-series
spectroscopy (Sect. 4) clearly show us that the geometry of the
magnetosphere is far from a pristine axisymmetric dipole, but instead
probably consists of many azimuthally distributed funnels of accretion,
curved by rotation and varying in position relative to the stellar surface
on the timescale of a few stellar rotation periods.  Furthermore, the
temperature of the magnetosphere and the mass accretion rate are degenerate
quantities in the models, with a higher temperature magnetosphere producing
more line flux for the same accretion rate. This means that brown dwarf
models require a much higher accretion stream temperature than those of
CTTSs in order to produce the observed line flux, and although the
temperature is grossly constrained by the line broadening (which may
preclude lower temperature streams) the thermal structure of the accretion
streams is still a problem. Despite these uncertainties, and in defense of
the BD models, it should be noted that the low accretion rates derived are
consistent with both the lack of optical veiling ({\em Muzerolle et al.},
2003a) and the strength of the Ca\,{\sc ii} $\lambda 8662$ line ({\em
  Mohanty et al.}, 2005).

Current models do not match the line core particularly well, which is often
attributed to a break down of the Sobolev approximation; co-moving frame
calculations (which are many orders of magnitude more expensive
computationally) may be required. An additional problem with current RT
modelling is the reliance on fitting a single profile -- current studies
have almost always been limited to H$\alpha$ -- one that rarely shows an
IPC profile ({\em Edwards et al.}, 1994; {\em Reipurth et al.}, 1996), is
vulnerable to contamination by outflows (e.g., {\em Alencar et al.}, 2005)
and may be significantly spatially extended ({\em Takami et al.}, 2003).
Even in modelling a single line, it is fair to say that the
state-of-the-art is some way short of line profile fitting; the best fits
reported in the literature may match the observation in terms of peak
intensity, equivalent width, or in the line wings, but are rarely
convincing reproductions of the observations in detail. Only by
simultaneously fitting several lines may one have confidence in the models,
particularly if those lines share a common upper/lower level (H$\alpha$ and
Pa$\beta$ for example). Although such observations are in the literature
(e.g., {\em Edwards et al.}, 1994; {\em Folha and Emerson}, 2001) their
usefulness is marginalized by the likely presence of significant
variability between the epochs of the observations at the different
wavelengths: simultaneous observations of a wide range of spectral
diagnostics are required. Despite the caveats described above, line profile
modelling remains a useful (and in the BD case the only) route to the mass
accretion rate, and there is real hope that the current factor~$\sim 5$
uncertainties in mass accretion rates derived from RT modelling may be
significantly reduced in the future.

\bigskip

\section{\textbf{OBSERVATIONAL EVIDENCE FOR MAGNETOSPHERIC ACCRETION}}
\bigskip

Observations seem to globally support the magnetospheric accretion
concept in CTTSs, which includes the presence of strong stellar
magnetic fields, the existence of an inner magnetospheric cavity of a
few stellar radii, magnetic accretion columns filled with free falling
plasma, and accretion shocks at the surface of the stars. While this
section summarizes the observational signatures of magnetospheric
accretion in T Tauri stars, there is some evidence that the general
picture applies to a much wider range of mass, from young brown dwarfs
({\em Muzerolle et al.}, 2005; {\em Mohanty et al.}, 2005) to Herbig
Ae-Be stars ({\em Muzerolle et al.}, 2004; {\em Calvet et al.}, 2004;
{\em Sorelli et al.}, 1996).

In recent years, the rapidly growing number of detections of strong
stellar magnetic fields at the surface of young stars seem to put the
magnetospheric accretion scenario on a robust ground (see Sect. 2). As
expected from the models, given the typical mass accretion rates
($10^{-9}$ to $10^{-7}$ M$_\odot$ yr$^{-1}$, {\em Gullbring et al.},
1998) and magnetic field strengths ($2$ to $3$ kG, {\em Valenti and
Johns-Krull}, 2004) obtained from the observations, circumstellar disk
inner holes of about 3-9 R$_*$ are required to explain the observed
line widths of the CO fundamental emission, that likely come from gas
in Keplerian rotation in the circumstellar disk of CTTSs ({\em Najita
et al.}, 2003).  There has also been evidence for accretion columns
through the common occurrence of inverse P Cygni profiles with
redshifted absorptions reaching several hundred km s$^{-1}$, which
indicates that gas is accreted onto the star from a distance of a few
stellar radii ({\em Edwards et al.}, 1994).

Accretion shocks are inferred from the rotational modulation of light
curves by bright surface spots ({\em Bouvier et al.}, 1995) and modelling
of the light curves suggests hot spots covering about one percent of the
stellar surface.  The theoretical prediction of accretion shocks and its
associated hot excess emission are also supported by accretion shock models
that successfully reproduce the observed spectral energy distributions of
optical and UV excesses ({\em Calvet and Gullbring}, 1998; {\em Ardila and
  Basri}, 2000; {\em Gullbring et al.}, 2000). In these models, the
spectral energy distribution of the excess emission is explained as a
combination of optically thick emission from the heated photosphere below
the shock and optically thin emission from the preshock and postshock
regions. {\em Gullbring et al.} (2000) also showed that the high mass
accretion rate CTTSs have accretion columns with similar values of energy
flux as the moderate to low mass accretion rate CTTSs, but their accretion
columns cover a larger fraction of the stellar surface (filling factors
ranging from less than 1\% for low accretors to more than 10\% for the high
one). A similar trend was observed by {\em Ardila and Basri} (2000) who
found, from the study of the variability of IUE spectra of BP Tau, that the
higher the mass accretion rate, the bigger the hot spot size.
%Magnetospheric accretion models further yield reliable mass accretion rate
%estimates that can be used to calibrate other accretion diagnosis that are
%easier to measure than the UV excess, such as the de-redenned U-band excess
%and the IR luminosity of emission lines like Pa$\beta$ and Br$\gamma$.  
%The
%determination of trustful mass accretion rates is actually of key
%importance in the accretion theory to help us understand the evolution of
%both disk and star.

Statistical correlations between line fluxes and mass accretion rates
predicted by magnetospheric accretion models have also been reported
for emission lines in a broad spectral range, from the UV to the near-IR
({\em Johns-Krull et al.}, 2000; {\em Beristain et al.}, 2001; {\em Alencar
  and Basri}, 2000; {\em Muzerolle et al.}, 2001; {\em Folha and Emerson},
2001).  However, in recent years, a number of observational results
indicate that the idealized steady-state axisymmetric dipolar
magnetospheric accretion models cannot account for many observed
characteristics of CTTSs.

\begin{figure}[t]
\epsscale{1.0} \plotone{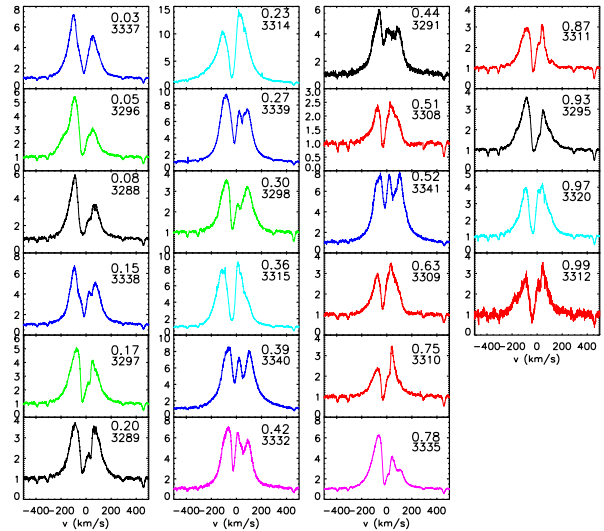}
\caption{The rotational modulation of the H$\alpha$ line profile of
  the CTTS AA Tau (8.2d period). Line profiles are ordered by
  increasing rotational phase (top panel number) at different Julian
  dates (bottom panel number). Note the development of a high velocity
  redshifted absorption component in the profile from phase 0.39 to
  0.52, when the funnel flow is seen against the hot accretion shock
  (from {\em Bouvier et al.}, in prep.).}
\label{aaha}
\end{figure}

Recent studies showed that accreting systems present strikingly large
veiling variability in the near-IR ({\em Eiroa et al.}, 2002; {\em Barsony
  et al.}, 2005), pointing to observational evidence for time variable
accretion in the inner disk. Moreover, the near-IR veiling measured in
CTTSs is often larger than predicted by standard disk models ({\em Folha
  and Emerson}, 1999; {\em Johns-Krull and Valenti}, 2001). This suggests
that the inner disk structure is significantly modified by its interaction
with an inclined stellar magnetosphere and thus departs from a flat disk
geometry. Alternatively, a ``puffed'' inner disk rim could result from the
irradiation of the inner disk by the central star and accretion shock ({\em
  Natta et al.} 2001; {\em
  Muzerolle et al.}, 2003b). In mildly accreting T Tauri stars, the dust
sublimation radius computed from irradiation models is predicted to lie
close to the corotation radius (3-9 R$_*$ $\simeq$ 0.03-0.08 AU) though
direct interferometric measurements tend to indicate larger values
(0.08-0.2 AU, {\em Akeson et al.}, 2005).

Observational evidence for an inner disk warp has been reported by {\em
  Bouvier et al.} (1999, 2003) for AA Tau, as expected from the interaction
between the disk and an {\em inclined} stellar magnetosphere (see Sect.~5).
Inclined magnetospheres are also necessary to explain the observed periodic
variations over a rotational timescale in the emission line and veiling
fluxes of a few CTTSs ({\em Johns and Basri}, 1995; {\em Petrov et al.},
1996; 2001; {\em Bouvier et al.}, 1999; {\em Batalha et al.}, 2002). These
are expected to arise from the variations of the projected funnel and shock
geometry as the star rotates. An example can be seen in Fig.~\ref{aaha}
that shows the periodic modulation of the H$\alpha$ line profile of the
CTTS AA Tau as the system rotates, with the development of a high velocity
redshifted absorption component when the funnel flow is seen against the
hot accretion shock. Sometimes, however, multiple periods are observed in
the line flux variability and their relationship to stellar rotation is not
always clear (e.g.,  {\em Alencar and Batalha}, 2002; {\em Oliveira et al.},
2000).  The expected correlation between the line flux from the accretion
columns, and the continuum excess flux from the accretion shock is not
always present either ({\em Ardila and Basri}, 2000; {\em Batalha et al.},
2002), and the correlations predicted by static {\em dipolar}
magnetospheric accretion models are generally not seen ({\em Johns-Krull
  and Gafford}, 2002).

Winds are generally expected to be seen as forbidden emission lines or the
blueshifted absorption components of permitted emission lines. Some
permitted emission line profiles of high-mass accretion rate CTTSs,
however, do not always look like the ones calculated with magnetospheric
accretion models and this could be in part due to a strong wind
contribution to the emission profiles, given the high optical depth of the
wind in these cases ({\em Muzerolle et al.}, 2001; {\em Alencar et al.},
2005).  Accretion powered hot winds originating at or close to the stellar
surface have recently been proposed to exist in CTTSs with high mass
accretion rates ({\em Edwards et al.}, 2003). These winds are inferred from
the observations of P Cygni profiles of the He I line (10780 \AA) that
present blueshifted absorptions which extend up to -400 km/s.  {\em Matt
  and Pudritz} (2005) have argued that such stellar winds can extract a
significant amount of the star's angular momentum, thus helping regulate
the spin of CTTSs. Turbulence could also be important and help explain the
very wide ($\pm$ 500 km s$^{-1}$) emission line profiles commonly observed
in Balmer and MgII UV lines ({\em Ardila et al.}, 2002).

Synoptic studies of different CTTSs highlighted the dynamical aspect of the
accretion/ejection processes, which only recently has begun to be studied
theoretically by numerical simulations (see Sect.~5). The accretion process
appears to be time dependent on several timescales, from hours for
non-steady accretion ({\em Gullbring et al.}, 1996; {\em Alencar and Batalha}, 
2002; {\em Stempels and Piskunov}, 2002; {\em Bouvier et al.}, 2003) to 
weeks for rotational modulation ({\em Smith et al.}, 1999; {\em Johns 
and Basri}, 1995; {\em Petrov et al.}, 2001), and from months for global 
instabilities of the magnetospheric structure ({\em Bouvier et al.}, 2003) 
to years for EXor and FUor eruptions (e.g., {\em Reipurth and Aspin}, 2004; 
{\em Herbig}, 1989).

One reason for such a variability could come from the interaction between
the stellar magnetosphere and the inner accretion disk.  In general,
magnetospheric accretion models assume that the circumstellar disk is
truncated close to the corotation radius and that field lines threading the
disk corotate with the star.  However, many field lines should interact
with the disk in regions where the star and the disk rotate differentially.
Possible evidence has been reported for differential rotation between the
star and the inner disk ({\em Oliveira et al.}, 2000) through the presence
of an observed time delay of a few hours between the appearance of high
velocity redshifted absorption components in line profiles formed in
different regions of the accretion columns.  This was interpreted as
resulting from the crossing of an azimuthally twisted accretion column on
the line of sight.  Another possible evidence for twisted magnetic field
lines by differential rotation leading to reconnection events has been
proposed by {\em Montmerle et al.}  (2000) for the embedded protostellar
source YLW 15, based on the observations of quasi-periodic X-ray flaring.
A third possible evidence was reported by {\em Bouvier et al.} (2003) for
the CTTS AA Tau. On timescales of the order of a month, they observed
significant variations in the line and continuum excess flux, indicative of
a smoothly varying mass accretion rate onto the star.  At the same time,
they found a tight correlation between the radial velocity of the
blueshifted (outflow) and redshifted (inflow) absorption components in the
H$\alpha$ emission line profile. This correlation provides support for a
physical connection between time dependent inflow and outflow in CTTSs.
{\em Bouvier et al.}  (2003) interpreted the flux and radial velocity
variations in the framework of magnetospheric inflation cycles due to
differential rotation between the star and the inner disk, as observed in
recent numerical simulations (see Sect.~5).  The periodicity of such
instabilities, as predicted by numerical models, is yet to be tested
observationally and will require monitoring campaigns of chosen CTTSs
lasting for several months.

\bigskip

\section{\textbf{NUMERICAL SIMULATIONS OF MAGNETOSPHERIC ACCRETION}}
\bigskip

\begin{figure*}[t]
\epsscale{1.3} \plotone{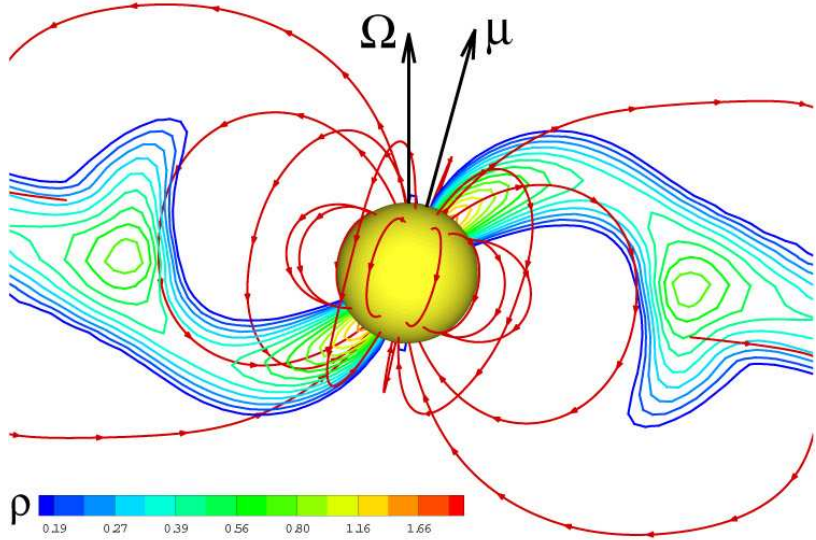} \caption{A slice of the funnel
stream obtained in 3D simulations for an inclined dipole
($\Theta=15^\circ$). The contour lines show density levels, from the
minimum (dark) to the maximum (light). The corona above the disk has a
low-density but is not shown.  The thick lines depict magnetic field
lines (from {\em Romanova et al.}, 2004a).}
\label{fig1}
\end{figure*}

Significant progress has been made in recent years in the numerical
modeling of magnetospheric accretion onto a rotating star with a dipolar
magnetic field.  One of the main problems is to find adequate initial
conditions which do not destroy the disk in first few rotations of the star
and do not influence the simulations thereafter. In particular, one must
deal with the initial discontinuity of the magnetic field between the disk
and the corona, which usually leads to significant magnetic braking of the
disk matter and artificially fast accretion onto the star on a dynamical
time-scale.  Specific quasi-equilibrium initial conditions were developed,
which helped to overcome this difficulty ({\em Romanova et al.}, 2002).  In
axisymmetric (2D) simulations, the matter of the disk accretes inward
slowly, on a viscous time-scale as expected in actual stellar disks. The
rate of accretion is regulated by a viscous torque incorporated into the
numerical code through the $\alpha$ prescription, with typically
$\alpha_{\rm v}=0.01-0.03$.

Simulations have shown that the accretion disk is disrupted by the stellar
magnetosphere at the magnetospheric or truncation radius $R_T$, where the
gas pressure in the disk is comparable to the magnetic pressure,
$P_{ram}=B^2/{8\pi}$ (see Sect.~2). In this region matter is lifted above
the disk plane due to the pressure force and falls onto the stellar surface
supersonically along the field lines, forming funnel flows ({\em Romanova
  et al.}, 2002). The location of the inner disk radius oscillates as a
result of accumulation and reconnection of the magnetic flux at this
boundary, which blocks or ``permits" accretion (see discussion of this
issue below), thus leading to non-steady accretion through the funnel
flows. Nevertheless, simulations have shown that the funnel flow is a
quasi-stationary feature during at least $50-80$ rotation periods of the
disk at the truncation radius, $P_0$, and recent simulations with improved
numerical schemes indicate that this structure survives for more than 
$1,000~P_0$ ({\em Long et al.}, 2005).  Axisymmetric simulations thus
confirmed the theoretical ideas regarding the structure of the accretion
flow around magnetized CTTSs. As a next step, similar initial conditions
were applied to full 3D simulations of disk accretion onto a star with an
{\em inclined} dipole, a challenging problem which required the development
of new numerical methods (e.g., the ``inflated cube" grid, cf.  {\em
  Koldoba et al.}, 2002; {\em Romanova et al.}, 2003, 2004a).  Simulations
have shown that the disk is disrupted at the truncation radius $R_T$, as in
the axisymmetric case, but the magnetospheric flow to the star is more
complex. Matter flows around the magnetosphere and falls onto the stellar
surface supersonically.  The magnetospheric structure varies depending on
the misalignment angle of the dipole, but settles into a quasi-stationary
state after a few $P_0$, as demonstrated by recent simulations run up to
$40~P_0$ ({\em Kulkarni and Romanova}, 2005).  In both, 2D and 3D
simulations the fluxes of matter and angular momentum to or from the star
vary in time, however they are smooth on average. This average value is
determined by the properties of the accretion disk.

Numerical simulation studies have shown that a star may either spin up,
spin down or be in rotational equilibrium when the net torque on the star
vanishes. Detailed investigation of the rotational equilibrium state has
shown that the rotation of the star is then {\it locked} at an angular
velocity $\Omega_{eq}$ which is smaller by a factor of $\sim 0.67 - 0.83$
than the angular velocity at the truncation radius ({\em Long et al.},
2005). The corresponding ``equilibrium" corotation radius $R_{CO}\approx
(1.3-1.5)~R_T$ is close to that predicted theoretically (e.g., {\em Ghosh
  and Lamb}, 1978,1979b; {\em K\"onigl}, 1991). Recently, the disk-locking
paradigm was challenged by a number of authors (e.g., {\em Agapitou and
  Papaloizou}, 2000; {\em Matt and Pudritz}, 2004, 2005). The skepticism
was based on the fact that the magnetic field lines connecting the star to
the disk may inflate and open, (e.g., {\em Aly and Kuijpers}, 1990; {\em
  Lovelace et al.}, 1995; {\em Bardou}, 1999; {\em Uzdensky et al.}, 2002),
resulting in a significant decrease of angular momentum transport between
the star and the disk. Such an opening of field lines was observed in a
number of simulations (e.g., {\em Miller and Stone}, 1997; {\em Romanova et
  al.}, 1998; {\em Fendt and Elstner}, 2000). Several factors, however,
tend to restore an efficient disk-star connection. One of them is that the
inflated field lines have a tendency to reconnect and close again ({\em
  Uzdensky et al.}, 2002). Furthermore, there is always a region of closed
field lines connecting the inner regions of the disk with the
magnetosphere, which provides angular momentum transport between the disk
and the star (e.g., {\em Pringle and Rees}, 1972; {\em Ghosh and Lamb},
1979b). This is the region where matter accretes through funnel flows and
efficiently transports angular momentum to or from the star. This torque
tends to bring a star in co-rotation with the inner regions of the disk.
There is always, however, a smaller but noticeable negative torque either
connected with the region $r > R_{CO}$ ({\em Ghosh and Lamb}, 1978,1979b),
if the field lines are closed in this region, or associated to a wind which
carries angular momentum out along the open field lines connecting the star
to a low-density corona.  Simulations have shown that the wind is
magnetically-dominated ({\em Long et al.}, 2005; {\em Romanova et al.},
2005), though the possibility of an accretion-driven {\it stellar} wind has
also been discussed ({\em Matt and Pudritz}, 2005). The spin-down through
magnetic winds was proposed earlier by {\em Tout and Pringle} (1992). Both
torques are negative so that in rotational equilibrium a star rotates {\it
  slower} than the inner disk.  Thus, the result is similar to the one
predicted earlier theoretically, though the physics of the spin-down
contribution may be different.  Axisymmetric simulations of the {\it fast
  rotating} CTTSs have shown that they efficiently spin-down through both
disk-magnetosphere interaction and magnetic winds ({\em Romanova et al.},
2005; {\em Ustyugova et al.}, 2006). For instance, it was shown that a CTTS
with an initial period $P=1$ d spins down to the typically observed periods
of about a week in less that $10^6$ yr.

\begin{figure}[t]
  \epsscale{1.} \plotone{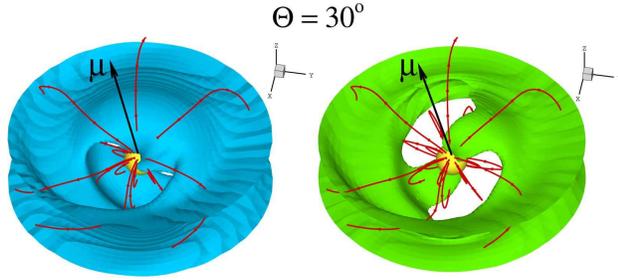} \caption{3D simulations show that
    matter accretes onto the star through narrow, high density streams
    (right panel) surrounded by lower density funnel flows that blanket
    nearly the whole magnetosphere (left panel). 
  }
\label{fig2}
\end{figure}

Three-dimensional (3D) simulations of disk accretion onto a star with a
misaligned dipolar magnetic field have shown that at the non-zero
misalignment angle $\Theta$, where $\Theta$ is an angle between the
magnetic moment $\bf{\mu_*}$ and the rotational axis $\bf{\Omega_*}$ of the
star (with the disk axis aligned with $\bf{\Omega_*}$), matter typically
accretes in two and, in some cases, in several streams ({\em Koldoba et
  al.}, 2002; {\em Romanova et al.}, 2003, 2004a). Figure~\ref{fig1} shows
a slice of the magnetospheric stream at $\Theta=15^\circ$. The density and
pressure of the flow increase towards the star as a result of the
convergence of the flow. They are also larger in the central regions of the
funnel streams. Thus, the structure of the magnetospheric flow depends on
the density. The high density part is channeled in narrow funnel streams,
while the low density part is wider, with accreting matter blanketting the
magnetosphere nearly completely ({\em Romanova et al.}, 2003, see
Figure~\ref{fig2}). The spectral lines which form in the funnel streams are
redshifted or blueshifted depending on the angle $\Theta$ and viewing angle
$i$ and their strength is modulated by the rotation of the star.

\begin{figure}[t]
\epsscale{1.} \plotone{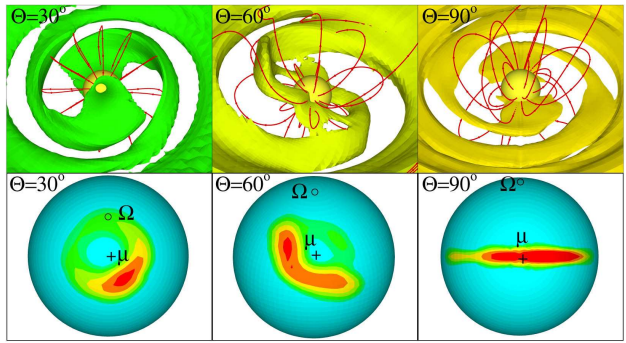} \caption{Top panels: matter flow
close to the star at different misalignment angles $\Theta$. Bottom
panels: the shape of the corresponding hot spots. Darker regions 
correspond to larger density. (From {\em Romanova et al.}, 2004a).}
\label{fig3}
\end{figure}

Matter in the funnel flows falls onto the star's surface and forms
{\it hot spots}. The shape of the spots and the distribution of
different parameters (density, velocity, pressure) in the spots
reflect those in the cross-section of the funnel streams ({\em
Romanova et al.}, 2004a). Figure~\ref{fig3} shows an example of
magnetospheric flows and hot spots at different $\Theta$. At
relatively small angles, $\Theta\lesssim 30^{\circ}$, the spots have
the shape of a bow, while at very large angles, $\Theta \gtrsim
60^\circ$, they have a shape of a bar crossing the surface of the star
near the magnetic pole. The density, velocity and pressure are the
largest in the central regions of the spots and decrease outward (see
Figure~\ref{fig3}). The temperature also increases towards the center
of the spots because the kinetic energy flux is the largest there. The
rotation of the star with surface hot spots leads to variability with
one or two peaks per period depending on $\Theta$ and $i$. The two
peaks are typical for larger $\Theta$ and $i$. The position of the
funnel streams on the star is determined by both the angular velocity
of the star and that of the inner radius of the disk.  In the
rotational equilibrium state, the funnel flows usually settle in a
particular ``favorite'' position. However, if the accretion rate
changes slightly, say, increases, then the truncation radius decreases
accordingly and the angular velocity at the foot-point of the funnel
stream on the disk is larger. As a result, the other end of the stream
at the surface of the star changes its position by a small
amount. Thus, the location of the spots ``wobbles" around an
equilibrium position depending on the accretion rate ({\em Romanova et
al.}, 2004a). The variation of the accretion rate also changes the
size and the brightness of the spots.

The disk-magnetosphere interaction leads to the thickening of the inner
regions of the disk which eases the lifting of matter to the funnel flow.
Matter typically accumulates near the closed magnetosphere forming a denser
ring ({\em Romanova et al.}, 2002) which brakes into a spiral structure in
case of misaligned dipole ({\em Romanova et al.}, 2003, 2004a). Typically,
two trailing spiral arms are obtained (see Figure~\ref{fig3}). 3D
simulations have also shown that when accretion occurs onto a tilted
dipole, the inner regions of the disk are slightly warped. This results
from the tendency of disk material to flow along the magnetic equator of
the misaligned dipole ({\em Romanova et al.}, 2003). Such a warping is
observed for medium misalignment angles, $30^\circ < \Theta < 60^\circ$.
Disk warping in the opposite direction (towards magnetic axis of the
dipole) was predicted theoretically when the disk is strongly diamagnetic
({\em Aly}, 1980; {\em Lipunov and Shakura}, 1980; {\em Lai}, 1999). The
warping of the inner disk and the formation of a spiral structure in the
accretion flow may possibly be at the origin of the observed variability of
some CTTSs ({\em Terquem and Papaloizou}, 2000; {\em Bouvier et al.},
2003).

Progress has also been made in the modeling of outflows from the vicinity
of the magnetized stars. Such outflows may occur from the
disk-magnetosphere boundary ({\em Shu et al.}, 1994), from the disk ({\em
  Blandford and Payne}, 1982; {\em Pudritz and Norman}, 1986; {\em Lovelace
  et al.}, 1991; {\em Lovelace et al.}, 1995; {\em Casse and Ferreira},
2000; {\em Pudritz et al.}, 2006), or from the star ({\em Matt and
  Pudritz}, 2005). Magneto-centrifugally driven outflows were first
investigated in pioneering short-term simulations by {\em Hayashi et al.}
(1996) and {\em Miller and Stone} (1997) and later in longer-term
simulations with a fixed disk ({\em Ouyed and Pudritz}, 1997; {\em Romanova
  et al.}, 1997; {\em Ustyugova et al.}, 1999; {\em Krasnopolsky et al.},
1999; {\em Fendt and Elstner}, 2000). Simulations including feedback on the
inner disk have shown that the process of the disk-magnetosphere
interaction is non-stationary: the inner radius of the disk oscillates, and
matter accretes to the star and outflows quasi-periodically ({\em Goodson
  et al.}, 1997, 1999; {\em Hirose et al.}, 1997; {\em Matt et al.}, 2002;
{\em Kato et al.}, 2004; {\em Romanova et al.}, 2004b; {\em Von Rekowski
  and Brandenburg}, 2004; {\em Romanova et al.}, 2005), as predicted by
{\em Aly and Kuijpers} (1990).  The characteristic timescale of variability
is determined by a number of factors, including the time-scale of diffusive
penetration of the inner disk matter through the external regions of the
magnetosphere ({\em Goodson and Winglee}, 1999).  It was earlier suggested
that reconnection of the magnetic flux at the disk-magnetosphere boundary
may lead to X-ray flares in CTTSs ({\em Hayashi et al.}, 1996; {\em
  Feigelson and Montmerle}, 1999) and evidence for very large flaring
structures has been recently reported by {\em Favata et al.} (2005).

So far simulations were done for a dipolar magnetic field.
Observations suggest a non-dipolar magnetic field near the stellar
surface (see Sect. 2, and also, e.g., {\em Safier}, 1998; {\em
Kravtsova and Lamzin}, 2003; {\em Lamzin}, 2003; {\em Smirnov et al.},
2005). If the dipole component dominates on the large scale, many
properties of magnetospheric accretion will be similar to those
described above, including the structure of the funnel streams and
their physical properties. However, the multipolar component will
probably control the flow near the stellar surface, possibly affecting
the shape and the number of hot spots. Simulations of accretion to a
star with a multipolar magnetic field are more complicated, and
should be done in the future.

\bigskip

\section{\textbf{CONCLUSIONS}} 
\bigskip

Recent magnetic field measurements in T Tauri stars support the view
that the accretion flow from the inner disk onto the star is
magnetically controlled. While typical values of 2.5~kG are obtained
for photospheric fields, it also appears that the field topology is
likely complex on the small-scales (R$\leq$R$_\star$), while on the
larger scale (R$>>$R$_\star$) a globally more organized but weaker
($\sim$0.1 kG) magnetic component dominates. This structure is thought
to interact with the inner disk to yield magnetically-channeled
accretion onto the star. Observational evidence for magnetospheric
accretion in classical T Tauri star is robust (inner disk truncation,
hot spots, line profiles) and the rotational modulation of
accretion/ejection diagnostics observed in some systems suggests that
the stellar magnetosphere is moderately inclined relative to the
star's rotational axis. Realistic 3D numerical models have capitalized
on the observational evidence to demonstrate that many properties of
accreting T Tauri stars could be interpreted in the framework of
magnetically-controlled accretion.  One of the most conspicuous
properties of young stars is their extreme variability on timescales
ranging from hours to months, which can sometimes be traced to
instabilities or quasi-periodic phenomena associated to the magnetic
star-disk interaction.

Much work remains to be done, however, before reaching a complete
understanding of this highly dynamical and time variable process.
Numerical simulations still have to incorporate field geometries more
complex than a tilted dipole, e.g., the superposition of a large-scale
dipolar or quadrupolar field with multipolar fields at smaller
scales. The modeling of emission line profiles now starts to combine
radiative transfer computations in both accretion funnel flows and
associated mass loss flows (disk winds, stellar winds), which indeed
appears necessary to account for the large variety of line profiles
exhibited by CTTSs. These models also have to address the strong line
profile variability which occurs on a timescale ranging from hours to
weeks in accreting T Tauri stars. These foreseen developments must be
driven by intense monitoring of typical CTTSs on all timescales from
hours to years, which combines photometry, spectroscopy and polarimetry
in various wavelength domains.  This will provide strong constraints
on the origin of the variability of the various components of the
star-disk interaction process (e.g., inner disk in the near-IR, funnel
flows in emission lines, hot spots in the optical or UV, magnetic
reconnections in X-rays, etc.).

The implications of the dynamical nature of magnetospheric accretion in
CTTSs are plentiful and remain to be fully explored. They range from the
evolution of stellar angular momentum during the pre-main sequence phase
(e.g., {\em Agapitou and Papaloizou}, 2000), the origin of inflow/ouflow
short term variability (e.g.,  {\em Woitas et al.}, 2002; {\em Lopez-Martin
  et al.}, 2003), the modeling of the near infrared veiling of CTTSs and of
its variations, both of which will be affected by a non planar and time
variable inner disk structure (e.g., {\em Carpenter et al.}, 2001; {\em
  Eiroa et al.}, 2002), and possibly the halting of planet migration close
to the star ({\em Lin et al.}, 1996).

\bigskip
%\newpage

\textbf{ Acknowledgments.} SA acknowledges financial support from CNPq
through grant 201228/2004-1. Work of MMR was supported by the NASA
grants NAG5-13060, NAG5-13220, and by the NSF grants AST-0307817 and
AST-0507760.

\centerline\textbf{ REFERENCES}
\bigskip
\parskip=0pt
{\small
\baselineskip=11pt

\refs Agapitou V. and Papaloizou J.~C.~B. (2000) {\em
Mon. Not. R. Astron. Soc., 317}, 273-288.

\refs Akeson R.~L., Boden A.~F., Monnier J.~D., Millan-Gabet R.,
Beichman C., et al. (2005) {\em Astrophys. J., 635}, 1173-1181.

\refs Alencar S.~H.~P. and  Basri G. (2000) {\em Astron. J., 119},
1881-1900. 

\refs Alencar S.~H.~P. and Batalha C. (2002) {\em Astrophys. J., 571}, 378-393.

\refs Alencar S.~H.~P., Basri G., Hartmann L., and Calvet
N. (2005) {\em Astron. Astrophys., 440}, 595-608.

\refs Aly J.~J. (1980) {\em Astron. Astrophys., 86}, 192-197.

\refs Aly J.~J. and Kuijpers J. (1990) {\em Astron. Astrophys.,
227}, 473-482.

\refs Andr\'e P. (1987) In {\em Protostars and Molecular Clouds} (T. Montmerle
and C. Bertout, eds.), pp.143-187. CEA, Saclay.

\refs Ardila D.~R. and Basri G. (2000) {\em Astrophys. J., 539}, 834-846. 

\refs Ardila D.~R., Basri G., Walter F.~M., Valenti J.~A., 
and Johns-Krull C.~M. (2002) {\em Astrophys. J., 567}, 1013-1027. 

\refs Bardou A. (1999) {\em Mon. Not. R. Astron. Soc., 306},
669-674.

\refs Barsony M., Ressler M.~E., and Marsh K.~A. (2005)
{\em Astrophys. J., 630}, 381-399. 

\refs Basri G. and Bertout C. (1989) {\em Astrophys. J., 341}, 340-358.

\refs Basri G., Marcy G.~W., and Valenti J.~A. (1992)
{\em Astrophys. J., 390}, 622-633.

\refs Batalha C., Batalha N.~M., Alencar S.~H.~P., Lopes D.~F., 
and Duarte E.~S. (2002) {\em Astrophys. J., 580}, 343-357. 

\refs Beristain G., Edwards S., and Kwan J. (2001) {\em Astrophys. J., 551}, 1037-1064. 

\refs Blandford R.~D. and Payne D.~G. (1982) {\em Mon. Not. R. Astron.
Soc., 199}, 883-903.

\refs Borra E.~F., Edwards G., and Mayor M. (1984) {\em Astrophys. J., 284}, 
211-222. 

\refs Bouvier J., Chelli A., Allain S., Carrasco L., Costero
R., et al. (1999) {\it Astron. Astrophys., 349}, 619-635. 

\refs Bouvier J., Grankin K.~N., Alencar S.~H.~P., Dougados C., 
Fern{\`a}ndez M., et al. (2003) {\em Astron. Astrophys.,
409}, 169-192.

\refs Bouvier J., Cabrit S., Fenandez M., Martin E.~L., and Matthews
J.~M. (1993) {\it Astron. Astrophys., 272}, 176-206.

\refs Bouvier J., Covino E., Kovo O., Mart\'in E.~L., Matthews J.~M.,
et al. (1995) {\it Astron. Astrophys., 299},
89-107.

\refs Brown D.~N. and Landstreet J.~D. (1981) {\em
Astrophys. J., 246}, 899-904.

\refs Calvet N. and Gullbring E.\ (1998) {\em Astrophys. J., 509},
802-818. 

\refs Calvet N., Muzerolle J., Brice\~no C., Fernandez J., Hartmann
L., et al. (2004) {\it Astron. J., 128}, 1294-1318.

\refs Camenzind M. (1990) {\em Rev. Mex. Astron. Astrofis., 3}, 234-265.

\refs Carpenter J.~M., Hillenbrand L.~A., and Skrutskie M.~F. (2001) {\em Astron.
  J., 121}, 3160-3190.

\refs Casse F. and Ferreira J. (2000) {\it Astron. Astrophys. 361}, 1178-1190

\refs Collier Cameron A.~C. and Campbell C.~G. (1993) {\it
Astron. Astrophys., 274}, 309-318.

\refs Daou A.~G., Johns--Krull C.~M., and Valenti J.~A.
(2006) {\it Astron. J., 131}, 520-526.

\refs Donati J.-F., Semel M., Carter B.~D., Rees D.~E.,
and Collier Cameron A. (1997) {\em Mon. Not. R. Astron. Soc., 291}, 658-682.

\refs Edwards S., Hartigan P., Ghandour L., and Andrulis 
C. (1994) {\it Astron. J., 108}, 1056-1070. 

\refs Edwards S., Fischer W., Kwan J., Hillenbrand L., and 
Dupree A.~K. (2003) {\em Astrophys. J., 599}, L41-L44.

\refs Eiroa, C., Oudmaijer R.~D., Davies J.~K., de Winter D., Garzón F., et al.
(2002) {\em Astron. Astrophys., 384}, 1038-1049.

\refs Favata F., Flaccomio E., Reale F., Micela G., Sciortino S., et al.
(2005) {\em Astrophys. J. Suppl., 160}, 469-502.

\refs Feigelson E.~D. and Montmerle T. (1999) {\em Ann. Rev.
Astron. Astrophys., 37}, 363-408.

\refs Fendt C. and Elstner D. (2000) {\em Astron. Astrophys.,
363}, 208-222.

\refs Folha D.~F.~M. and Emerson J.~P. (1999) {\it
Astron. Astrophys., 352}, 517-531.

\refs Folha D.~F.~M. and Emerson J.~P. (2001) {\it
Astron. Astrophys., 365}, 90-109.

\refs Ghosh P. and Lamb F.~K. (1978) {\em Astrophys. J.,
223}, L83-L87.

\refs Ghosh P. and Lamb F.~K. (1979a) {\it Astrophys. J.,
232}, 259-276.

\refs Ghosh P. and Lamb F.~K. (1979b) {\em Astrophys. J., 234},
296-316. 

\refs Goodson A.~P. and Winglee R.~M. (1999) {\em
Astrophys. J., 524}, 159-168.

\refs Goodson A.~P., Winglee R.~ M., and B\"ohm K.-H. (1997) {\em
Astrophys. J., 489}, 199-209.

\refs Goodson A.~P., B\"ohm K.-H. and Winglee R.~M. (1999) {\em
Astrophys. J., 524}, 142-158.

\refs Guenther E.~W., Lehmann H., Emerson J.~P., and
Staude J. (1999) {\it Astron. Astrophys., 341}, 768-783.

\refs Gullbring E., Barwig H., Chen P.~S., Gahm G.~F., and Bao 
M.~X. (1996) {\em Astron. Astrophys., 307}, 791-802. 

\refs Gullbring E., Hartmann L., Brice\~no C., and
Calvet N. (1998) {\em Astrophys. J., 492}, 323-341.

\refs Gullbring E., Calvet N., Muzerolle J., and Hartmann L. (2000) 
{\em Astrophys. J., 544}, 927-932. 

\refs Hartigan P., Edwards S., and Ghandour L. (1995) {\em Astrophys. J., 452},
736-768.

\refs Hartmann L., Hewett R., and Calvet N.  (1994) {\em Astrophys. J., 426}, 669-687.

\refs Hayashi M.~R., Shibata K., and Matsumoto R. (1996) {\em
Astrophys. J., 468}, L37-L40.

\refs Herbig G.~H. (1989) In {\em Low Mass Star Formation and Pre-main Sequence
  Objects} (B. Reipurth, ed.), pp.233-246. ESO, Garching.

\refs Herbst W., Bailer-Jones C.~A.~L., Mundt R.,
Meisenheimer K., and Wackermann R. (2002) {\it Astron. Astrophys.,
396}, 513-532.

\refs Hirose S., Uchida Y., Shibata K., and Matsumoto R.
(1997) {\em Pub. Astron. Soc. Jap., 49}, 193-205.

\refs Jardine M., Collier Cameron A., and Donati J.-F.
(2002) {\em Mon. Not. R. Astron. Soc., 333}, 339-346.

\refs Johns C.~M. and Basri G. (1995) {\em Astrophys. J., 449}, 341-364. 

\refs Johns--Krull C.~M. and Gafford A.~D. (2002) {\em
Astrophys. J., 573}, 685-698.

\refs Johns--Krull C.~M. and Valenti J.~A. (2001) {\em
Astrophys. J., 561}, 1060-1073.

\refs Johns--Krull C.~M., Valenti J.~A., Hatzes A.~P., and
Kanaan A. (1999a) {\em Astrophys. J., 510}, L41-L44.

\refs Johns--Krull C.~M., Valenti J.~A., and Koresko
C. (1999b) {\em Astrophys. J., 516}, 900-915.

\refs Johns-Krull C.~M., Valenti J.~A., and Linsky J.~L. (2000) 
{\em Astrophys. J., 539}, 815-833. 

\refs Johns--Krull C.~M., Valenti J.~A., Saar S.~H., and Hatzes A.~P.
(2001) In {\em Magnetic Fields Across the Hertzsprung-Russell Diagram}
(G. Mathys et al., eds.), pp. 527-532. ASP Conf. Series, San Francisco. 

\refs Johns--Krull C.~M., Valenti J.~A., and Saar S.~H.
(2004) {\em Astrophys. J., 617}, 1204-1215.

\refs Johnstone R.~M. and Penston M.~V. (1986) {\em
Mon. Not. R. Astron. Soc., 219}, 927-941.

\refs Johnstone R.~M. and Penston M.~V. (1987) {\em
Mon. Not. R. Astron. Soc., 227}, 797-800.

\refs Kato Y., Hayashi M.~R. and Matsumoto R. (2004) {\em
Astrophys. J., 600}, 338-342.

\refs Koide S., Shibata K., and Kudoh T. (1999) {\em Astrophys. J., 522},
727-752.

\refs Koldoba A.~V., Romanova M.~M., Ustyugova G.~V., and Lovelace
R.~V.~E. (2002) {\em Astrophys. J., 576}, L53 -L56.

\refs K\"onigl. A. (1991) {\em Astrophys. J., 370},
L39-L43.

\refs Krasnopolsky R., Li Z.-Y., and Blandford R.
(1999) {\em Astrophys. J., 526}, 631-642.

\refs Kravtsova A.~S. and Lamzin S.~A. (2003) {\em Astron.
Lett., 29}, 612-620.

{\refs Kulkarni A.~K. and Romanova M.~M. (2005) {\em
Astrophys. J., 633}, 349-357.

\refs Kurosawa R., Harries T.~J., and Symington N.~H. (2006) {\em
Mon. Not. R. Astron. Soc.}, submitted

\refs Lai D. (1999) {\em Astrophys. J., 524}, 1030-1047.

\refs Lamzin S.~A. (2003) {\em Astron. Reports, 47},
498-510.

\refs Lawson W.~A., Lyo A.-R., and Muzerolle J. (2004) {\em Mon.
Not. R. Astron. Soc., 351}, L39-L43.

\refs Lin D.~N.~C., Bodenheimer P., and Richardson D.~C. (1996) {\em Nature,
  380}, 606-607.

\refs Lipunov V.~M. and Shakura N.~I. (1980)
 {\em  Soviet Astronomy Letters, 6}, 14-17.

\refs Long M., Romanova M.~M., and Lovelace R.~V.~E. (2005)
{\em Astrophys. J., 634}, 1214-1222.

\refs L{\'o}pez-Mart{\'{\i}}n L., Cabrit S., and Dougados C. (2003) {\em Astron.
  Astrophys., 405}, L1-L4.

\refs Lovelace R.~V.~E., Berk H.~L., and Contopoulos J. (1991) {\em
Astrophys. J., 379}, 696-705.

\refs Lovelace R.~V.~E., Romanova M.~M., and Bisnovatyi-Kogan G.~S.
(1995) {\em Mon. Not. R. Astron. Soc., 275}, 244-254.

\refs Martin S.~C., (1996) {\em Astrophys. J., 470}, 537-550.

\refs Matt S. and Pudritz R.~E. (2004) {\em Astrophys. J., 607},
L43-L46.

\refs Matt S. and Pudritz R.~E. (2005) {\em Astrophys.
J., 632}, L135-L138.

\refs Matt S., Goodson A.~P., Winglee R.~M., and B\"ohm K.-H.
(2002) {\em Astrophys. J., 574}, 232-245.

\refs M{\'e}nard F. and Bertout C. (1999) In {\em The Origin of Stars and
  Planetary Systems} (C.~J. Lada and N.~D. Kylafis, eds), pp.341. Kluwer
Academic Publishers.

\refs Miller K.~A. and Stone J.~M. (1997) {\em Astrophys. J.,
489}, 890-902.

\refs Mohanty S.~M., Jayawardhana R., and Basri G. (2005) {\em
Astrophys. J., 626}, 498-522.

\refs Montmerle T., Koch-Miramond L., Falgarone E., and Grindlay J.~E. (1983)
{\em Astrophys. J., 269}, 182-201.

\refs Montmerle T., Grosso N., Tsuboi Y., and Koyama K. (2000) 
{\em Astrophys. J., 532}, 1097-1110.

\refs Muzerolle J., Calvet N., and Hartmann L. (2001) {\em
Astrophys. J., 550}, 944-61.

\refs Muzerolle J., Hillenbrand L., Calvet N., Brice\~no C., and Hartmann
L. (2003a) {\em Astrophys. J., 592}, 266-281.

\refs Muzerolle J., Calvet N., Hartmann L., and D'Alessio P. (2003b)
{\em Astrophys. J., 597}, L149-L152.

\refs Muzerolle J., D'Alessio P., Calvet N., and Hartmann L. (2004)
{\em Astrophys. J., 617}, 406-417.

\refs Muzerolle J., Luhman K.~L., Brice\~no C., Hartmann L.,
and Calvet N. (2005) {\em Astrophys. J., 625}, 906-912.

\refs Najita J., Carr J.~S. and Mathieu R.~D. (2003) {\em
Astrophys. J., 589}, 931-952.

\refs Natta A., Prusti T., Neri R., Wooden D., Grinin V.~P., and Mannings
V. (2001) {\it Astron.  Astrophys., 371}, 186-197

\refs Oliveira J.~M. Foing, B.~H., van Loon J.~T., and Unruh Y.~C. (2000) 
{\it Astron. Astrophys., 362}, 615-627.

\refs Ostriker E.~C. and Shu F.~H. (1995) {\em
Astrophys. J., 447}, 813-828.

\refs Ouyed R. and Pudritz R.~E. (1997) {\em Astrophys.
J., 482}, 712-732.

\refs {\bf ud} Petit P., Donati J.-F., Wade G.~A., Landstreet J.~D.,
Bagnulo S., et al. (2004) {\em Mon. Not. R. Astron. Soc., 348}, 1175-1190.

\refs Petrov P.~P., Gullbring E., Ilyin I., Gahm G.~F., Tuominen I.,
et al. (1996) {\it Astron. Astrophys., 314}, 821-834.

\refs Petrov P.~P., Gahm G.~F., Gameiro J.~F., Duemmler R., Ilyin
I.~V., et al.  (2001) {\it Astron. Astrophys., 369}, 993-1008.

\refs Pringle J.~E. and Rees M.~J. (1972) {\em Astron.
Astrophys., 21}, 1P-9P.

\refs Pudritz R.~E. and Norman C.~A. (1986) {\em
Astrophys. J., 301}, 571-586.

\refs Pudritz R.~E., Rogers C.~S., and Ouyed R. (2006) {\em Mon. Not. R. Astron.
  Soc., 365}, 1131-1148.

\refs Reipurth B. and Aspin C. (2004) {\em Astrophys. J., 608}, L65-L68.

\refs Reipurth B., Pedrosa A., and Lago M.~T.~V.~T. (1996) {\em
Astron. Astrophys. Suppl., 120}, 229-256.

\refs Romanova M.~M., Ustyugova G.~V., Koldoba A.~V.,
Chechetkin V.~M., and Lovelace R.~V.~E. (1997) {\em Astrophys. J.,
482}, 708-711.

\refs Romanova M.~M., Ustyugova G.~V., Koldoba A.~V., Chechetkin
V.~M., and Lovelace R.~V.~E. (1998) {\em Astrophys. J., 500},
703-713.

\refs Romanova M.~M., Ustyugova G.~V., Koldoba A.~V., and Lovelace
R.~V.~E. (2002) {\em Astrophys. J., 578}, 420-438.

\refs Romanova M.~M., Ustyugova G.~V., Koldoba A.~V., Wick J.~V.,
and Lovelace R.~V.~E. (2003) {\em Astrophys. J., 595}, 1009-1031.

\refs Romanova M.~M., Ustyugova G.~V., Koldoba A.~V., and Lovelace
R.~V.~E. (2004a) {\em Astrophys. J., 610}, 920-932.

\refs Romanova M.~M., Ustyugova G.~V., Koldoba A.~V., and Lovelace
R.~V.~E. (2004b) {\em Astrophys. J., 616}, L151 -L154.

\refs Romanova M.~M., Ustyugova G.~V., Koldoba A.~V., and Lovelace
R.~V.~E. (2005) {\em Astrophys. J., 635},
L165-L168.

\refs Safier P.~N. (1998) {\em Astrophys. J., 494}, 336-341.

\refs Shu F., Najita J., Ostriker E., Wilkin F., Ruden S., and
Lizano S. (1994) {\em Astrophys. J., 429}, 781-796.

\refs Smirnov D.~A., Lamzin S.~A., Fabrika S.~N., and
Valyavin G.~G. (2003) {\em Astron. Astrophys., 401}, 1057-1061.

\refs Smirnov D.~A., Lamzin S.~A., Fabrika S.~N., and
Chuntonov G.~A. (2004) {\em Astron. Lett., 30}, 456-460.

\refs Smirnov D.~A., Romanova M.~M., and Lamzin S.~A. (2005)
{\em Astron. Lett., 31}, 335-339.

\refs Smith K.~W., Lewis G.~F., Bonnell I.~A., Bunclark P.~S., and 
Emerson J.~P. (1999) {\em Mon. Not. R. Astron. Soc., 304}, 367-388. 

\refs Smith K., Pestalozzi M., G{\"u}del M., Conway J., and Benz A.~O. (2003)
{\em Astron. Astrophys., 406}, 957-967.

\refs Sorelli C., Grinin V.~P., and Natta A. (1996) {\it Astron.
  Astrophys., 309}, 155-162.

\refs Stempels H.~C. and Piskunov N. (2002) {\it
Astron. Astrophys., 391}, 595-608.

\refs Symington N.~H., Harries T.~J., and Kurosawa R. (2005a) {\em
Mon. Not. R. Astron. Soc., 356}, 1489-1500.

\refs Symington N.~H., Harries T.~J., Kurosawa R., and
Naylor T.  (2005b) {\em Mon. Not. R. Astron. Soc., 358}, 977-984.

\refs Takami M., Bailey J., and Chrysostomou A. (2003) {\em
Astron. Astrophys., 397}, 675-984.

\refs Terquem C. and Papaloizou J.~C.~B. (2000) {\em
Astron. Astrophys., 360}, 1031-1042.

\refs Tout C.~A. and Pringle J.~E. (1992) {\em Mon. Not. R. Astron. Soc., 256},
269-276.

\refs Ustyugova G.~V., Koldoba A.~V., Romanova M.~M., and Lovelace
R.~V.~E. (1999) {\em Astrophys. J., 516}, 221-235.

\refs Ustyugova, G.V., Koldoba, A.V., Romanova, M.M., and Lovelace,
R.V.E. (2006). {\em Astrophys. J.}, in press

\refs Uzdensky D.~A., K\"onigl A., and Litwin C. (2002) {\em
Astrophys. J., 565}, 1191-1204.

\refs Valenti J.~A. and Johns-Krull C.~M. (2004)
{\em Astrophys. Sp. Sci. Ser., 292}, 619-629. 

\refs Valenti J.~A., Basri G., and Johns C.~M. (1993)
{\it Astron. J., 106}, 2024-2050.

\refs Vink J.~S., Drew J.~E., Harries T.~J., Oudmaijer R.~D.,
and Unruh Y. (2005a) {\em Mon. Not. R. Astron. Soc., 359}, 1049-1064.

\refs Vink J.~S., Harries T.~J., and Drew J.~E. (2005b) {\em
Astron. Astrophys., 430}, 213-222.

\refs Vogt S.~S. (1980) {\em Astrophys. J., 240}, 567-584. 

\refs von Rekowski B. and Brandenburg A. (2004) {\em Astron.
Astrophys., 420}, 17-32.

\refs Warner B. (2004) {\em Publ. Astron. Soc. Pac., 116}, 115-132

\refs Woitas J., Ray T.~P., Bacciotti F., Davis C.~J., and Eisl{\"o}ffel J.
(2002) {\em Astrophys. J., 580}, 336-342.

\refs Yang H., Johns--Krull C.~M., and Valenti J.~A.
(2005) {\em Astrophys. J., 635}, 466-475.

\
\end{document}